\documentclass[runningheads]{llncs}

\newif\iflong\longtrue

\usepackage{tikz}
\usepackage{ltl}
\usepackage{xspace}
\usepackage{todonotes}
\usepackage[inline]{enumitem}
\usepackage{wrapfig}
\usepackage{cite}
\usepackage{subcaption}
\usepackage{stmaryrd}
\usepackage{listings}
\usepackage{multirow}

\allowdisplaybreaks

\usepackage[ruled,vlined,linesnumbered]{algorithm2e}
\SetKwInOut{Input}{Input}
\SetKwInOut{Output}{Output}
\SetKwProg{Fn}{Function}{}{}
\SetNlSkip{0.3ex} 
\SetAlgoSkip{}
\usepackage{hyperref}
\hypersetup{%
  colorlinks=true,           
  allcolors=blue!70!black,   
  pdfstartview=Fit,          
  breaklinks=true,
  pdfauthor={B. Abraham, E. Bartocci, B. Bonakdarpout, O. Dobe},
  pdftitle={Probabilistic Hyperproperties with Nondeterminism}}

\usetikzlibrary{arrows,backgrounds,decorations,decorations.pathmorphing,
positioning,fit,automata,shapes,snakes,patterns,plotmarks,calc,trees}

\hypersetup{breaklinks=true}
\usepackage{soul}

\newcommand{\erika}[2]{#2}

\begin{document}

\newcommand{\plus}{+}

\newcommand{\sep}{ \ \ | \ \ }
\newcommand{\suchthat}{\,|\,}

\newcommand{\taskone}{1\xspace}
\newcommand{\tasktwo}{2\xspace}
\newcommand{\taskthree}{3\xspace}
\newcommand{\taskfour}{4\xspace}

\newcommand{\z}{\cellcolor{black}}
\newcommand{\x}{\cellcolor{lightgray}}

\newcommand{\xpath}{\pi}
\newcommand{\Paths}[2]{\mathit{Paths}_{#1}^{#2}}
\newcommand{\fPaths}[2]{\mathit{fPaths}_{#1}^{#2}}

\newcommand{\dom}{\mathit{dom}}
\newcommand{\dtmc}{\mathcal{D}}
\newcommand{\mdp}{\mathcal{M}}
\newcommand{\LTL}{\textsf{\small LTL}\xspace}
\newcommand{\CTL}{\textsf{\small CTL}\xspace}
\newcommand{\CTLstar}{\textsf{\small CTL$^*$}\xspace}
\newcommand{\PCTL}{\textsf{\small PCTL}\xspace}
\newcommand{\PCTLstar}{\textsf{\small PCTL$^*$}\xspace}
\newcommand{\HyperPCTL}{\textsf{\small HyperPCTL}\xspace}
\newcommand{\PHL}{\textsf{\small PHL}\xspace}
\newcommand{\NHyperPCTL}{\textsf{\small HyperPCTL}\xspace}
\newcommand{\HyperLTL}{\textsf{\small HyperLTL}\xspace}
\newcommand{\HyperCTL}{\textsf{\small HyperCTL}\xspace}
\newcommand{\HyperCTLstar}{\textsf{\small HyperCTL$^*$}\xspace}

\newcommand{\lang}{\mathcal{L}}
\newcommand{\Inf}{\mathsf{Inf}}
\newcommand{\pba}{\mathcal{P}}
\newcommand{\alphabet}{\mathrm{\Sigma}}
\newcommand{\state}{s}
\newcommand{\vstate}{\vec{\state}}
\newcommand{\states}{S}
\newcommand{\hstate}{\hat{\state}}
\newcommand{\hstateof}[2]{\hat{\state}_{#1}(#2)}
\newcommand{\hstates}{\hat{\states}}

\newcommand{\bstate}{\bar{\state}}
\newcommand{\bstateof}[2]{\bar{\state}_{#1}(#2)}
\newcommand{\bstates}{\bar{\states}}

\newcommand{\map}{\mathsf{map}}

\newcommand{\scheduler}{\sigma}
\newcommand{\sched}{\scheduler}
\newcommand{\minscheduler}{\scheduler_{*}}
\newcommand{\maxscheduler}{\scheduler^{*}}
\newcommand{\vscheduler}{\vec{\scheduler}}
\newcommand{\schedulers}[1]{\Sigma^{#1}}

\newcommand{\hscheduler}{\hat{\scheduler}}
\newcommand{\hschedulers}{\hat{\schedulers{}}}

\newcommand{\bscheduler}{\bar{\scheduler}}
\newcommand{\bschedulers}{\hat{\schedulers{}}}

\newcommand{\btruth}[2]{\textit{holds}_{#1,#2}}
\newcommand{\ptruth}[2]{\textit{prob}_{#1,#2}}
\newcommand{\btoi}[2]{\textit{holdsToInt}_{#1,#2}}

\newcommand{\formula}{\varphi}
\newcommand{\qform}{\varphi^{q}}
\newcommand{\nqform}{\varphi^{\textit{nq}}}
\newcommand{\prform}{\varphi^{\textit{pr}}}
\newcommand{\pathform}{\varphi^{\textit{path}}}

\newcommand{\propof}[2]{#1_{#2}}

\newcommand{\Trace}{\mathsf{Traces}}
\newcommand{\trace}{t}
\newcommand{\qtrace}{\eta}
\newcommand{\sform}{\mathrm{\Phi}}
\newcommand{\pform}{\varphi}

\newcommand{\action}{\alpha}
\newcommand{\altaction}{\beta}

\newcommand{\ins}{\textit{inst}}
\newcommand{\conc}{\circ}

\newcommand{\modes}{Q}
\newcommand{\mode}{q}
\newcommand{\modef}{\textit{mode}}

\renewcommand{\qed}{$~\blacksquare$}
\newcommand{\naturals}{\mathbb{N}_{>0}}
\newcommand{\naturalszero}{\mathbb{N}_{\geq 0}}

\newcommand{\F}{\LTLdiamond}
\newcommand{\G}{\LTLsquare}
\newcommand{\U}{\,\mathcal U\,}
\newcommand{\X}{\LTLcircle}
\newcommand{\Waitfor}{\,\mathcal W\,}
\newcommand{\suffix}[2]{#1[#2,\infty]}

\newcommand{\AP}{\mathsf{AP}}
\newcommand{\Next}{\X}
\newcommand{\Finally}{\F}
\newcommand{\Globally}{\G}
\newcommand{\V}{\mathcal{V}}

\newcommand{\pr}{\mathbb{P}}
\renewcommand{\Pr}{\mathrm{Pr}}
\newcommand{\Cyl}{\textit{Cyl}}
\newcommand{\parallelsum}{\mathbin{\|}}
\newcommand{\emptyword}{\epsilon}

\newcommand{\init}{\mathit{\iota_{init}}}
\renewcommand{\P}{\mathbf{P}}
\newcommand{\Act}{\mathit{Act}}
\newcommand{\act}{\mathit{act}}
\newcommand{\supp}{\mathit{supp}}
\newcommand{\start}{\mathit{init}}
\newcommand{\tru}{\mathtt{true}}
\newcommand{\fals}{\mathtt{false}}
\newcommand{\quant}{\mathbb{Q}}

\newcommand{\dbsim}{\mathit{dbSim}}
\newcommand{\res}{\mathit{res}}
\newcommand{\qout}{\mathit{qOut}}
\newcommand{\env}{\mathit{env}}
\newcommand{\fail}{\mathit{fail}}

\newcommand{\comp}[1]{\textsf{\small #1}}

\newcommand{\sigmakp}{$\mathsf{\Sigma^p_k}$\comp{-complete}\xspace}
\newcommand{\pikp}{$\mathsf{\Pi^p_k}$\comp{-complete}\xspace}

\newcommand\donotshow[1]{}

\definecolor{mGreen}{rgb}{0,0.6,0}
\definecolor{mGray}{rgb}{0.5,0.5,0.5}
\definecolor{mPurple}{rgb}{0.58,0,0.82}
\definecolor{backgroundColour}{rgb}{0.95,0.95,0.92}

\lstdefinestyle{CStyle}{
    backgroundcolor=\color{backgroundColour},   
    commentstyle=\color{mGreen},
    keywordstyle=\color{magenta},
    numberstyle=\tiny\color{mGray},
    stringstyle=\color{mPurple},
    basicstyle=\footnotesize,
    breakatwhitespace=false,         
    breaklines=true,                 
    captionpos=b,                    
    keepspaces=true,                 
    numbers=left,                    
    numbersep=2pt,                  
    showspaces=false,                
    showstringspaces=false,
    showtabs=false,                  
    tabsize=2,
    language=C
}

\newcommand{\borzoo}[1]{\textcolor{green}{#1}}

\title{Probabilistic Hyperproperties with Nondeterminism\thanks{This research 
has been partially supported by the United States NSF SaTC Award 181338, by the Vienna Science and Technology Fund ProbInG Grant ICT19-018 and by the DFG Research and Training Group UnRAVeL.
The 
order of
 authors is alphabetical and all authors made equal contribution.}} 

\author{Erika {\'{A}}brah{\'{a}}m \inst{1} \and
	Ezio Bartocci \inst{2} \and
	Borzoo Bonakdarpour\inst{3} \and
	Oyendrila Dobe \inst{3}}


\institute{RWTH-Aachen, Germany, 
\email{abraham@informatik.rwth-aachen.de} \and
Technische Universit\"at Wien, Austria, \email{ezio.bartocci@tuwien.ac.at} \and 
Michigan State University, USA, \email{\{borzoo,dobeoyen\}@msu.edu}}

\maketitle

\begin{abstract}

We study the problem of formalizing and checking
probabilistic hyperproperties for models that allow nondeterminism in
actions. We extend the temporal logic \HyperPCTL, which has been
previously introduced for discrete-time Markov chains, to enable the
specification of hyperproperties also for Markov decision
processes. We generalize \HyperPCTL by allowing explicit and
simultaneous quantification over schedulers and probabilistic
computation trees and show that it can express important quantitative
requirements in security and privacy. We show that \HyperPCTL model
checking over MDPs is in general undecidable for quantification over
probabilistic schedulers with memory, but restricting the domain to
memoryless non-probabilistic schedulers turns the model checking
problem decidable. Subsequently, we propose an SMT-based encoding for 
model checking this language and evaluate its performance.

\end{abstract}

\section{Introduction}
\label{sec:intro}

{\em Hyperproperties}~\cite{cs10} extend the conventional notion of {\em trace 
properties}~\cite{as85} from a set of traces to a set of sets of traces. In 
other words, a hyperproperty stipulates a {\em system} property and not the 
property of just individual traces. It has been shown that many interesting 
requirements in computing systems are hyperproperties and cannot be 
expressed by trace properties. Examples include (1) a wide range of 
information-flow security policies such as {\em noninterference}~\cite{gm82} 
and {\em observational determinism}~\cite{zm03}, (2) sensitivity and robustness 
requirements in cyber-physical systems~\cite{wzbp19}, and consistency 
conditions such as {\em linearizability} in concurrent data 
structures~\cite{bss18}.

Hyperproperties can describe the requirements of probabilistic systems as well. 
They generally express probabilistic relations between multiple executions of a 
system. For example, in information-flow security, adding probabilities is 
motivated by establishing a connection between information theory and 
information flow across multiple traces. A prominent example is probabilistic 
schedulers that open up an opportunity for an attacker to set up a 
probabilistic covert channel. Or, {\em probabilistic causation} compares the 
probability of occurrence of an effect between scenarios where the cause is or 
is not present.

\begin{wrapfigure}{r}{4.2cm}
\centering
\vspace{-10mm}
\scalebox{.75}{
\begin{tikzpicture}

\node[draw,circle,text width=0.5cm] (s0) at (0,0) {};
\node at (0,0) {$s_0$};
\node[draw,circle,text width=0.5cm] (s1) at (2,0) {};
\node at (2,0) {$s_1$};
\node[draw,circle,text width=0.5cm] (s2) at (-1,-1) {};
\node at (-1,-1) {$s_2$};
\node[draw,circle,text width=0.5cm] (s3) at (1,-1) {};
\node at (1,-1) {$s_3$};
\node[draw,circle,text width=0.5cm] (s4) at (3,-1) {};
\node at (3,-1) {$s_4$};
\node[draw,circle,text width=0.5cm] (s5) at (0,-2) {};
\node at (0,-2) {$s_5$};
\node[draw,circle,text width=0.5cm] (s6) at (2,-2) {};
\node at (2,-2) {$s_6$};

\node[left of=s0,node distance=0.9cm] {$\{\textit{init}\}$};
\node[right of=s1,node distance=0.9cm] {$\{\textit{init}\}$};
\node[left of=s2,node distance=0.7cm] {$\{a\}$};
\node[right of=s6,node distance=0.7cm] {$\{a\}$};

\path[draw,-latex',thick] (s0) edge node[left,font=\scriptsize,near start] {$0.4$} (s2);
\path[draw,-latex',thick] (s0) edge node[right,font=\scriptsize,near start=5pt] {$0.2$} (s3);
\path[draw,-latex',thick] (s0) edge node[right,font=\scriptsize] {$0.4$} (s5);
\path[draw,-latex',thick] (s1) edge node[left,font=\scriptsize,near start] {$0.7$} (s3);
\path[draw,-latex',thick] (s1) edge node[right,font=\scriptsize,near start] {$0.3$} (s4);
\path[draw,-latex',thick] (s2) edge node[left,font=\scriptsize] {$1$} (s5);
\path[draw,-latex',thick] (s3) edge node[right,font=\scriptsize] {$0.8$} (s5);
\path[draw,-latex',thick] (s3) edge node[right,font=\scriptsize,near start] {$0.2$} (s6);
\path[draw,-latex',thick] (s4) edge node[left,font=\scriptsize,near start] {$1$} (s6);
\path[draw,-latex'] (s5) edge [loop below] node[right,font=\scriptsize] {$1$} (s5);
\path[draw,-latex'] (s6) edge [loop below] node[right,font=\scriptsize] {$1$} (s6);
\end{tikzpicture}
}
\vspace*{-5mm}
 \caption{Example DTMC.}
 \label{fig:dtmc}
\vspace*{-8mm}
\end{wrapfigure}
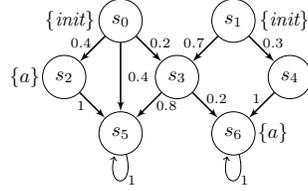
The state of the art on probabilistic hyperproperties has exclusively been 
studied in the context of discrete-time Markov chains (DTMCs). In~\cite{ab18}, 
we proposed the temporal logic \HyperPCTL, which extends \PCTL by allowing 
explicit and 
simultaneous quantification over computation trees. For example, the DTMC 
in Fig.~\ref{fig:dtmc} satisfies the following \HyperPCTL formula:
\begin{align}
 \label{eq:hpctl}
\psi = \forall \hstate.\forall \hstate'.  
\Big(\textit{init}_{\hstate}\wedge\textit{init}_{\hstate'}\Big) \Rightarrow
  \Big(\pr(\F a_{\hstate}) = \pr(\F a_{\hstate'})\Big)
\end{align}
which means that the probability of reaching proposition $a$ from any pair of 
states $\hstate$ and $\hstate'$ labeled by $\mathit{init}$ should be equal. 
Other works on probabilistic hyperproperties for DTMCs include parameter 
synthesis~\cite{abbd20} and statistical model checking~\cite{wnbp21,wzbp19}.

\begin{wrapfigure}{r}{5cm} 
\vspace{-8mm}
\centering
\scalebox{.75}{
\begin{tikzpicture}
  \node[draw,circle] (s0) at (-1,1) {$s_0$};
  \node[node distance=6ex,left of = s0] {\footnotesize $\{h{>}0\}$};

  \node[draw,circle] (s1) at (1,1) {$s_1$};
  \node[node distance=6ex,right of = s1] {\footnotesize $\{h{\leq}0\}$};

  \node[draw,circle] (s2) at (-2,-1.5) {$s_2$};
  \node[node distance=5.2ex,left of = s2] {\footnotesize $\{l{=}1\}$};

  \node[draw,circle] (s3) at (2,-1.5) {$s_3$};
  \node[node distance=5.2ex,right of = s3] {\footnotesize $\{l{=}2\}$};

  \node (n3) at (-2,-0.25) [circle,fill,inner sep=2pt]{};
  \node (n4) at (-0.7,0.15) [circle,fill,inner sep=2pt]{};
  \node (n1) at (0.7,0.15) [circle,fill,inner sep=2pt]{};
  \node (n2) at (2,-0.25) [circle,fill,inner sep=2pt]{};

  \node (n5) at (2,-2.5) [circle,fill,inner sep=2pt]{};
  \node (n6) at (-2,-2.5) [circle,fill,inner sep=2pt]{};

\draw (s0)-- node[left]{\footnotesize $\alpha$} (n3);
\draw[-latex'] (n3)-- node[left,pos=0.2,yshift=2pt]{\footnotesize $\frac{3}{4}$} 
(s2);
\draw[-latex'] (n3)-- node[right,pos=0.05,yshift=4pt]{\footnotesize 
$\frac{1}{4}$} (s3);

\draw (s0)-- node[right]{\footnotesize $\beta$} (n4);
\draw[-latex'] (n4)-- node[left,pos=0.1,yshift=4pt]{\footnotesize 
$\frac{1}{2}$} 
(s2);
\draw[-latex'] (n4)-- node[right,pos=0.05,yshift=4pt]{\footnotesize 
$\frac{1}{2}$} (s3);

\draw (s1)-- node[left]{\footnotesize $\alpha$} (n1);
\draw[-latex'] (n1)-- node[left,pos=0.05,yshift=4pt]{\footnotesize 
$\frac{2}{3}$} (s2);
\draw[-latex'] (n1)-- node[right,pos=0.1,yshift=4pt]{\footnotesize 
$\frac{1}{3}$} (s3);

\draw (s1)-- node[right]{\footnotesize $\beta$} (n2);
\draw[-latex'] (n2)-- node[left,pos=0.05,yshift=4pt]{\footnotesize 
$\frac{1}{2}$} (s2);
\draw[-latex'] (n2)-- node[right,pos=0.2,yshift=2pt]{\footnotesize 
$\frac{1}{2}$} (s3);


\path (s3) edge[->, bend right] node[left,pos=0.2,yshift=-2pt] {\footnotesize 
$\tau$} (n5);
\path (n5) edge[->, bend right] node[right,pos=0.2,yshift=3pt] {\footnotesize 
$1$} (s3);

\path (s2) edge[->, bend right] node[left,pos=0.2,yshift=-2pt] {\footnotesize 
$\tau$} (n6);
\path (n6) edge[->, bend right] node[right,pos=0.2,yshift=3pt] {\footnotesize 
$1$} (s2);


\end{tikzpicture}
}
    \caption{Example MDP.}
    \label{fig:mdp}
  \vspace{-7mm}
\end{wrapfigure} 
An important gap in the spectrum is verification of probabilistic 
hyperproperties with regard to models that allow {\em nondeterminism}, in 
particular, {\em Markov decision processes} (MDP). Nondeterminism plays a 
crucial role in many probabilistic systems. For instance, nondeterministic 
queries can be exploited in order to make targeted attacks to databases with 
private information~\cite{guarnieri2017}. To motivate the idea, consider the 
MDP in Fig.~\ref{fig:mdp}, where $h$ is a high secret and $l$ is a low 
publicly observable variable. To protect the secret, there should be no 
probabilistic dependencies between observations on the low variable $l$ and the
value of $h$. However, an attacker that chooses a scheduler that always takes 
action $\alpha$ from states $s_0$ and $s_1$ can learn whether or not $h \leq 0$ 
by observing the probability of obtaining $l=1$ (or $l=2$). On the other hand, 
a scheduler that always chooses action $\beta$, does not leak any information 
about the value of $h$. Thus, a natural question to ask is whether a certain 
property holds 
for all or some schedulers.

\begin{wrapfigure}[6]{r}{3.5cm}
\centering 
\vspace{-8mm}
\begin{tikzpicture}
  \node[draw,circle] (s0) at (0,0) {$s_0$};

  \node[draw,circle] (s1) at (1.5,0) {$s_1$};

  \path (s0) edge[->] node[above] {\footnotesize $\alpha\ 1$} (s1);
  \path (s0) edge [loop above] node {\footnotesize $\beta\ 1$} (s0);
 \path (s1) edge [loop above] node {\footnotesize $\alpha\ 1$} (s1);

\end{tikzpicture}
\vspace{-2mm}
\caption{Example~MDP.}
\label{fig:bscc}
\end{wrapfigure}
With the above motivation, in this paper, we focus on probabilistic
hyperproperties in the context of MDPs. Such hyperproperties
inherently need to consider different nondeterministic choices in
different executions, and naturally call for quantification over
schedulers.
There are several challenges to achieve this.
In general,
there are schedulers whose reachability probabilities cannot be
achieved by any memoryless non-probabilistic scheduler, and, hence
finding a scheduler is not reducible to checking non-probabilistic
memoryless schedulers, as it is done in \PCTL mode checking for
MDPs. Consider for example the MDP in
Fig.~\ref{fig:bscc}, for which we want to know whether there is a
scheduler such that the probability to reach $s_1$ from
$s_0$ equals $0.5$.  There are two non-probabilistic memoryless
schedulers, one choosing action $\alpha$ and the other, action $\beta$
in $s_0$. The first one is the maximal scheduler for which $s_1$ is reached with
probability $1$, and the second one is
the minimal scheduler leading to probability $0$.
However, the probability 
$0.5$ cannot be achieved by
any non-probabilistic scheduler. {\em Memoryless}
probabilistic schedulers can neither achieve probability $0.5$: if a
memoryless scheduler would take action $\alpha$ with any positive
probability, then the probability to reach $s_1$ is always $1$. The
only way to achieve the reachability probability $0.5$ (or any value
strictly between $0$ and $1$) is by a probabilistic scheduler with
memory, e.g., taking $\alpha$ and $\beta$ in $s_0$ with probabilities
$0.5$ each when this is the first step on a path, and $\beta$ with
probability $1$ otherwise.

Our contributions in this paper are as follows. We first extend the
temporal logic \HyperPCTL~\cite{ab18} to the context of MDPs. To this
end, we augment the syntax and semantics of \HyperPCTL to quantify
over schedulers and relate probabilistic computation trees for
different schedulers. For example, the following formula
generalizes~\eqref{eq:hpctl} by requiring that the respective property should hold for all
computation trees starting in any states $\hstate$ and $\hstate'$ of
the DMTC induced by any scheduler $\hscheduler$:
\begin{align*}
\forall \hscheduler.\forall \hstate(\hscheduler).\forall \hstate'(\hscheduler). 
\Big(\mathit{init}_{\hstate} \wedge \mathit{init}_{\hstate'} \Big) \, 
\Rightarrow \, 
&\Big(\pr (\F a_{\hstate}) = \pr (\F a_{\hstate'})\Big) 
\end{align*}

On the negative side, we show that the problem to check \HyperPCTL
properties for MDPs is in general undecidable. On the positive side,
we show that the problem becomes decidable when we restrict the
scheduler quantification domain to memoryless non-probabilistic
schedulers. We also show that this restricted problem is already 
\comp{NP-complete} (respectively, \comp{coNP-complete}) in the size of the 
given MDP for \HyperPCTL formulas with a single existential (respectively, 
universal) scheduler quantifier. Subsequently, we propose an SMT-based encoding 
to solve the restricted model checking problem. We have implemented our method 
and analyze it experimentally on three case studies: probabilistic scheduling 
attacks,  side-channel timing attacks, and probabilistic conformance (available 
at \url{https://github.com/oreohere/HyperOnMDP}).

It is important to note that the work in~\cite{dft20} (also published in
ATVA'20) independently addresses the problem under investigation in this paper.
The authors propose the temporal logic \PHL. Similar to \HyperPCTL, \PHL also
allows quantification over schedulers, but path quantification of the induced
DTMC is achieved by using \HyperCTLstar. Both papers show that the model
checking problem is undecidable for the respective logics. The difference,
however, is in our approaches to deal with the undecidability result, which
leads two complementary and orthogonal techniques. For both logics the problem 
is decidable for non-probabilistic memoryless schedulers. We provide an 
SMT-based verification procedure for \HyperPCTL for this class of schedulers.
The work in~\cite{dft20} presents two methods for proving and for refuting 
formulas from a fragment of \PHL for general memoryful schedulers.
The two papers offer disjoint case studies for evaluation.

\paragraph{Organization.} Preliminary concepts are discussed in 
Section~\ref{sec:pre}. We present the syntax and semantics of \HyperPCTL for 
MDPs and discuss its expressive power in Section~\ref{sec:hpctl}. 
Section~\ref{sec:app} is dedicated to the applications of \HyperPCTL. 
Sections~\ref{sec:memless} and~\ref{sec:results} present our results on 
memoryless non-probabilistic schedulers and their evaluation before concluding 
in Section~\ref{sec:concl}.
\iflong
All proofs are given in the Appendix.
\else
More technical details and all proofs appear in~\cite{abbd20-arxiv}.
\fi

%
%

\section{Preliminaries}
\label{sec:pre}

\subsection{Discrete-time Markov models}

\begin{definition}
\label{def:dtmc}
A \emph{discrete-time Markov chain (DTMC)} is a tuple $\dtmc {=} (\states, \P, 
\AP, L)$ with the following components:

\begin{itemize}[topsep=2pt]
\item $\states$ is a nonempty finite set of {\em states};

\item $\P : \states \times \states \rightarrow [0, 1]$ is a {\em transition 
probability function} with 
$\sum_{\state' \in \states} \P(\state, \state') =1$, for all $\state \in 
\states$;

\item $\AP$ is a finite set of {\em atomic propositions}, and 

\item $L : \states \rightarrow 2^{\AP}$ is a \emph{labeling function}.\hfill\qed

\end{itemize}

\end{definition}


Fig~\ref{fig:dtmc} shows a simple DTMC. An (\emph{infinite}) \emph{path of 
$\dtmc$} is an infinite sequence $\xpath =
\state_0\state_1\state_2\ldots \in \states^\omega$ of states with
$\P(\state_i, \state_{i+1}) > 0$, for all $i \geq 0$; we write
$\xpath[i]$ for $\state_i$.  Let $\Paths{\state}{\dtmc}$ denote the
set of all (infinite) paths of $\dtmc$ starting in $\state$, and
$\fPaths{\state}{\dtmc}$ denote the set of all non-empty finite prefixes of
paths from $\Paths{\state}{\dtmc}$, which we call \emph{finite
  paths}. For a finite path $\xpath =\state_0\ldots s_k \in 
\fPaths{\state_0}{\dtmc}$, $k\geq 0$, we define $|\xpath|=k$.
We will also use the notations $\Paths{}{\dtmc}=\cup_{s\in
  \states}\Paths{\state}{\dtmc}$ and $\fPaths{}{\dtmc}=\cup_{s\in
  \states}\fPaths{\state}{\dtmc}$. A state $t\in \states$ is
\emph{reachable} from a state $s\in \states$ in $\dtmc$ if there
exists a finite path in $\fPaths{s}{\dtmc}$ with last state
$t$; 
we use 
$\fPaths{\state,T}{\dtmc}$ to denote the set of all finite paths from $\fPaths{\state}{\dtmc}$ with last state in $T\subseteq S$. A state $s\in
\states$ is \emph{absorbing} if $\P(\state,\state)=1$.

The \emph{cylinder set} $\Cyl^{\dtmc}(\xpath)$ of a
finite path $\xpath\in \fPaths{\state}{\dtmc}$ is
the set of all infinite paths of $\dtmc$ with prefix $\xpath$. The
\emph{probability space for $\dtmc$ and state $\state\in\states$} is
$(\Paths{\state}{\dtmc},\{\cup_{\xpath\in R}\Cyl^{\dtmc}(\xpath)\suchthat
R\subseteq\fPaths{\state}{\dtmc}\},\Pr^{\dtmc}_{\state})$, where the
\emph{probability} of the cylinderset of $\xpath\in \fPaths{\state}{\dtmc}$ is
$\Pr^{\dtmc}_{\state}(\Cyl^{\dtmc}(\xpath))=\Pi_{i=1}^{|\xpath|}\P(\xpath[i{-}1]
,\xpath[i])$. 

Note that the cylinder sets of two finite paths starting in the same
state are either disjoint or one is contained in the other. According
to the definition of the probability spaces, the total probability for
a set of cylinder sets defined by the finite paths
$R\subseteq\fPaths{\state}{\dtmc}$ is $\Pr^{\dtmc}(R)=\sum_{\xpath\in
  R'}\Pr^{\dtmc}_{\state}(\xpath)$ with $R'=\{\xpath\in R\suchthat \textit{no
  $\xpath'\in R\setminus\{\xpath\}$ is a prefix of $\xpath$}\}$. To
improve readability, we sometimes omit the DTMC index $\dtmc$ in the
notations when it is clear from the context.

\iflong
Parallel composition formalizes simultaneous runs in different DTMCs.
\fi
\begin{definition}
The \emph{parallel composition} of two DTMCs $\dtmc_i = (\states_i,
\P_i, \AP_i, L_i)$, $i=1,2$, is the DTMC $\dtmc_1\times\dtmc_2=(\states, \P, \AP,
L)$ with the following components:

\begin{itemize}[topsep=2pt]
\item $\states=\states_1\times\states_2$;
    
\item $\P : \states \times \states \rightarrow [0, 1]$ with 
$\P((\state_1,\state_2),(\state_1',\state_2'))=\P_1(\state_1,\state_1')\cdot 
\P_2(\state_2,\state_2')$, for all states $(\state_1,\state_2), 
(\state_1',\state_2') \in \states$;

\item $\AP=\AP_1\cup\AP_2$, and

\item $L : \states \rightarrow 2^{\AP}$ with $L((\state_1,\state_2)) = 
L_1(\state_1)\cup L_2(\state_2)$.\hfill\qed

\end{itemize}
\end{definition}


\iflong
\medskip

\noindent Markov decision processes extend DTMCs with non-deterministic choices.
\fi

\begin{definition}
A {\em Markov decision process} (\emph{MDP}) is a tuple $\mdp = (\states, \Act, 
\P, \AP, L)$ with the following components:

\begin{itemize}[topsep=2pt]

\item $\states$ is a nonempty finite set of {\em states};

\item $\Act$ is a nonempty finite set of {\em actions};

\item $\P : \states \times \Act \times \states \rightarrow [0, 1]$ is
a {\em transition probability function} such that for all $s \in 
\states$ the \emph{set of enabled actions} in $s$
$\Act(\state)=\{\action\in\Act \suchthat \sum_{\state' \in \states} \P(\state, 
\action, \state') =1\}$
is not empty and
$\sum_{\state' \in \states} \P(\state, \action, \state')=0$
for all $\action \in \Act\setminus\Act(\state)$;

\item $\AP$ is a finite set of {\em atomic propositions}, and

\item $L : \states \rightarrow 2^{\AP}$ is 
a \emph{labeling function}.\hfill\qed
\end{itemize}

 \end{definition}

\noindent Fig.~\ref{fig:mdp} shows a simple MDP. Schedulers can be used to eliminate the non-determinism in MDPs, inducing DTMCs with well-defined probability spaces.

%

\begin{definition}
\label{def:scheduler}
A {\em scheduler} for an MDP $\mdp = (\states, \Act, \P, \AP, L)$ is a tuple
$\scheduler=(\modes, \act, \modef, \start)$, where

\begin{itemize}[topsep=2pt]
\item $\modes$ is a countable set of \emph{modes};

\item $\act: \modes\times\ \states \times \Act\rightarrow [0,1]$ is a function 
for which $\sum_{\action\in\Act(\state)}\act(\mode,\state,\action)=1$ and
$\sum_{\action\in\Act\setminus\Act(\state)}\act(\mode,\state,\action)=0$
for all $\state\in \states$ and $\mode\in \modes$;

\item $\modef: \modes\times\states \rightarrow \modes$ is a \emph{mode 
transition} function, and

\item $\start: \states \rightarrow \modes$ is a function selecting a
starting mode for each state of $\mdp$.\hfill\qed
\end{itemize}
\end{definition}
Let 
$\schedulers{\mdp}$ denote the set of all schedulers for the MDP $\mdp$. A scheduler is \emph{finite-memory} if $\modes$ is finite, \emph{memoryless} if 
$|\modes|=1$, and \emph{non-probabilistic} if $\act(\mode,\state,\action) \in 
\{0,1\}$ for all  $\mode\in \modes$, $s\in \states$ and $\action\in\Act$. 

\begin{definition}
\label{def:induce}
Assume an MDP $\mdp= (\states, \Act, \P, \AP, L)$ and a scheduler 
$\scheduler=(\modes, \act, \modef, \start)\in \schedulers{\mdp}$ for $\mdp$.
The \emph{DTMC induced 
by $\mdp$ and $\scheduler$} is defined as $\mdp^\scheduler = 
(\states^\scheduler, 
\P^\scheduler, \AP,
L^{\scheduler})$ with $\states^\scheduler = \modes\times\states$,
\[
\P^\scheduler((\mode,\state),(\mode',\state'))=\left\{\begin{array}{l@{\qquad}l}
\sum_{\action\in\Act(\state)} \act(\mode,\state,\action) \cdot 
\P(\state,\action,\state') & \textit{if } \mode'=\modef(\mode,\state)\\ 
0 & \textit{otherwise}
\end{array}\right.
\]
and $L^\scheduler(\mode,\state)=L(\state)$ for all $\state,\state'\in\states$ 
and all $\mode,\mode'\in
\modes$. \hfill\qed
\end{definition}
A state $\state'$ is \emph{reachable} from $s\in \states$ in MDP $\mdp$ is 
there exists a scheduler $\scheduler$ for $\mdp$ such that $\state'$ is 
reachable from $s$ in
$\mdp^\scheduler$. A state $s\in \states$ is \emph{absorbing} in
$\mdp$ if $s$ is absorbing in $\mdp^\scheduler$ for all schedulers
$\scheduler$ for $\mdp$.
We sometimes omit the MDP index $\mdp$ in the notations 
when it is clear from the context.

\section{HyperPCTL for MDPs}
\label{sec:hpctl}


\iflong
In this section we extend \HyperPCTL from \cite{ab18} for DTMCs, to argue also about non-determinism in MDPs.
\fi

\subsection{HyperPCTL Syntax}

{\NHyperPCTL (quantified) state formulas} $\qform$ are inductively defined 
as follows:
\[
\begin{array}{llll}
\textit{\small quantified formula} & \qform & ::= & \forall \hscheduler.\qform \sep \exists 
\hscheduler.\qform \sep \forall \hstateof{}{\hscheduler}.\qform \sep \exists 
\hstateof{}{\hscheduler}.\qform \sep \nqform 
\\
\textit{\small non-quantified formula}  & \nqform & ::= & \tru \sep \propof{a}{\hstate} \sep \nqform \wedge \nqform \sep \neg 
\nqform \sep \prform< \prform \\
\textit{\small probability expression} & \prform & ::= & \pr(\pathform) \sep  
f(\prform_1, \dots, \prform_k)\\
\textit{\small path formula} & \pathform & ::= & \Next \nqform \sep \nqform \, 
\U \, \nqform \sep \nqform \, \U^{[k_1,k_2]} \,\nqform 

\end{array}
\]
where $\hscheduler$ is a \emph{scheduler variable}\footnote{We use the notation $\hscheduler$ for scheduler variables and $\scheduler$ for schedulers, and analogously $\hstate$ for state variables and $\state$ for states.} from an infinite set 
$\hschedulers$, $\hstate$ is a \emph{state variable} from an infinite set 
$\hstates$, $\nqform$ is a quantifier-free state formula, $a \in \AP$ is an 
atomic proposition, $\prform$ is a \emph{probability expression}, $f: [0,1]^k 
\rightarrow \mathbb{R}$ are $k$-ary \erika{elementary functions to express 
operations over probabilities,}{arithmetic operators (binary addition, unary/binary subtraction, binary multiplication) over probabilities,} where constants are viewed as $0$-ary functions, 
and $\pathform$ is a {\em path formula}, such that $k_1 \leq k_2 \in 
\naturalszero$. The probability operator $\pr$ allows the usage of 
probabilities in arithmetic constraints and relations.

A \NHyperPCTL construct $\formula$ (probability expression $\prform$, state 
formula $\qform$, $\nqform$ or path formula $\pathform$) is \emph{well-formed} 
if each occurrence of any $\propof{a}{\hstate}$ with $a\in\AP$ and 
$\hstate\in\hstates$ is in the scope of a {\em state quantifier} for 
$\hstateof{}{\hscheduler}$ for some $\hscheduler\in\hschedulers$, and any 
quantifier for $\hstateof{}{\hscheduler}$ is in the scope of a {\em 
scheduler quantifier} for $\hscheduler$. We restrict ourselves to quantifying 
first the schedulers then the states, i.e., different state variables can share 
the same scheduler. One can consider also \emph{local} schedulers when 
different players cannot explicitly share the same scheduler, or in other 
words, each scheduler quantifier belongs to exactly one of the quantified 
states.

\emph{\NHyperPCTL formulas} are well-formed \NHyperPCTL state 
formulas, where we additionally allow standard syntactic sugar like $\fals = \neg 
\tru$, $\varphi_1 \vee \varphi_2 = \neg(\neg \varphi_1 \wedge \neg\varphi_2)$, 
$\F \varphi = \tru \, \U \, \varphi$, and $\pr(\G \varphi) = 1-\pr(\F \neg\varphi)$.
For example, the \NHyperPCTL state formula 
$\forall\hscheduler.\exists\hstateof{}{\hscheduler}.\pr(\Next 
\propof{a}{\hstate})<0.5$ is a \NHyperPCTL formula.
  The \NHyperPCTL state formula $\pr(\Next \propof{a}{\hstate}){<}0.5$ is not a 
\NHyperPCTL formula, but can be extended to such. The \NHyperPCTL state formula 
$\forall\hstateof{}{\hscheduler}.\exists\hscheduler.\pr(\Next 
\propof{a}{\hstate}){<}0.5$ is not a \NHyperPCTL formula, and it even cannot 
can be extended to such.

\subsection{HyperPCTL Semantics}

\iflong
The semantics of \NHyperPCTL is based on the $n$-ary {\em self-composition} 
of an MDP.
\fi


\begin{definition}
The \emph{n-ary self-composition} of an MDP $\mdp = (\states, \Act,
\P, \AP, L)$ for a sequence
$\vscheduler=(\scheduler_1,\ldots,\scheduler_n)\in({\schedulers{\mdp}})^n$
of schedulers for $\mdp$ is the DTMC parallel composition
$\mdp^{\vscheduler}=\mdp_1^{\scheduler_1}\times\ldots\times\mdp_n^{\scheduler_n}$,
where $\mdp_i^{\scheduler_i}$ is the DTMC induced by $\mdp_i$ and
$\scheduler_i$, and where $\mdp_i=(\states,\Act,\P,\AP_i,L_i)$ with
$\AP_i=\{\propof{a}{i}\suchthat a\in\AP\}$ and
$L_i(s)=\{\propof{a}{i}\suchthat a\in L(s)\}$, for all $s\in S$. \hfill \qed
\end{definition}
%

\NHyperPCTL state formulas are evaluated in the context of an MDP $\mdp = 
(\states, \Act, \P, \AP, L)$, a sequence
$\vscheduler=(\scheduler_1,\ldots,\scheduler_n)\in(\schedulers{\mdp})^n$
of schedulers, and a sequence 
$\vec{r}=((\mode_1,\state_1),\ldots,(\mode_n,\state_n))$ of $\mdp^{\vscheduler}$ 
states; we use $()$ to denote
the empty sequence (of any type) and $\conc$ for 
concatenation. Intuitively, these sequences store
instantiations for scheduler and state variables. 
The satisfaction of a \NHyperPCTL quantified formula by $\mdp$ is defined by
\[
\mdp \models \varphi \qquad 
\textit{iff} \qquad
\mdp,(),()\models\varphi\ .
\]
The semantics evaluates \NHyperPCTL formulas by structural recursion.  Let in 
the following $\quant, \quant',\ldots$ denote quantifiers from 
$\{\forall,\exists\}$. When instantiating $\quant\hscheduler.\varphi$ 
by a scheduler $\scheduler \in \schedulers{\mdp}$, we replace in  $\varphi$ 
each subformula $\quant'\hstateof{}{\hscheduler}.\varphi'$, that is not in the 
scope of a quantifier for $\hscheduler$ by 
$\quant'\hstateof{}{\scheduler}.\varphi'$, and denote the result by 
$\varphi[\hscheduler\leadsto\scheduler]$. For instantiating a state quantifier 
$\quant \hstateof{}{\scheduler}.\varphi$ by a state $\state$, we append 
$\scheduler=(\modes, \act, \modef, \start)$ and $(\start(\state),\state)$ 
at the end of the respective sequences, and replace each $\propof{a}{\hstate}$ 
in the scope of the given quantifier by $\propof{a}{\state}$, resulting in a 
formula that we denote by $\varphi[\hstate\leadsto s]$. To evaluate 
probability expressions, we use the $n$-ary self-composition of the MDP.  

Formally, 
the semantics judgment rules are as follows:
{\small
\[
\begin{array}{l@{\quad}c@{\quad}l}
\mdp,\vscheduler,\vec{r} \models \tru\\
\mdp,\vscheduler,\vec{r} \models \propof{a}{i} & \textit{iff} & a_i\in 
L^{\vscheduler}(\vec{r}) \\
\mdp,\vscheduler,\vec{r}  \models \varphi_1 \wedge \varphi_2 &
\textit{iff} & \mdp,\vscheduler,\vec{r} \models \varphi_1 \textit{ and } 
\mdp,\vscheduler,\vec{r}  \models \varphi_2 \\
\mdp,\vscheduler,\vec{r}  \models \neg \varphi & \textit{iff} &
\mdp,\vscheduler,\vec{r} \not \models \varphi \\
\mdp,\vscheduler,\vec{r}\models \forall \hscheduler.\varphi & 
\textit{iff} & \forall \scheduler \in \schedulers{\mdp}.\ 
\mdp,\vscheduler,\vec{r}\models \varphi[\hscheduler\leadsto\scheduler]\\
\mdp,\vscheduler,\vec{r} \models \exists \hscheduler.\varphi & 
\textit{iff} & \exists \scheduler \in \schedulers{\mdp}.\ 
\mdp,\vscheduler,\vec{r}\models \varphi[\hscheduler\leadsto\scheduler] \\
\mdp,\vscheduler,\vec{r} \models \forall \hstateof{}{\scheduler}.\varphi & 
\textit{iff} & 
\forall \state_{n\plus 1} \in \states.\ \mdp,\vscheduler\conc\scheduler,\vec{r}\conc(\start(\state_{n\plus 1}),\state_{n\plus 1}) \models 
\varphi[\hstate \leadsto \state_{n\plus 1}]\\
\mdp,\vscheduler,\vec{r} \models \exists \hstateof{}{\scheduler}.\varphi & 
\textit{iff} & 
\exists \state_{n\plus 1} \in \states.\ \mdp,\vscheduler\conc\scheduler,\vec{r}\conc(\start(\state_{n\plus 1}),\state_{n\plus 1}) \models 
\varphi[\hstate \leadsto \state_{n\plus 1}] \\
\mdp,\vscheduler,\vec{r} \models \prform_1 < \prform_2 &
\textit{iff} &
\llbracket \prform_1\rrbracket_{\mdp,\vscheduler,\vec{r}} < \llbracket 
\prform_2 \rrbracket_{\mdp,\vscheduler,\vec{r}}\\
\llbracket \pr(\varphi_{\textit{path}})\rrbracket_{\mdp,\vscheduler,\vec{r}} 
& = & \Pr^{\mdp^{\vscheduler}}\big(\{\xpath \in 
\Paths{\vec{r}}{\mdp^{\vscheduler}} 
\mid \mdp,\vscheduler,\xpath \models \varphi_{\textit{path}}\}\big) \\
\llbracket f(\prform_1, \dots \prform_k)\rrbracket_{\mdp,\vscheduler,\vec{r}} 
& = & f\big(\llbracket \prform_1\rrbracket_{\mdp,\vscheduler,\vec{r}} 
\dots,\ \llbracket \prform_k\rrbracket_{\mdp,\vscheduler,\vec{r}}\big)
\end{array}
\]
}

\noindent where $\mdp$ is an MDP; $n\in\mathbb{N}_{\geq 0}$ is non-negative integer;
$\vscheduler\in(\schedulers{\mdp})^n$; $\vec{r}$ is a state of
$\mdp^{\vscheduler}$; $a\in\AP$ is an atomic proposition and
$i\in\{1,\ldots,n\}$; $\varphi, \varphi_1,\varphi_2$ are \NHyperPCTL
state formulas;
$\scheduler=(\modes, \act, \modef, \start)\in\schedulers{\mdp}$ is a
scheduler for $\mdp$; $\prform_1\cdots\prform_k$ are probability
expressions, and $\varphi_{\textit{path}}$ is a \NHyperPCTL path
formula whose satisfaction relation is as follows:
{\small
\[
\begin{array}{l@{\ }c@{\ \ }l}
\mdp,\vscheduler,\xpath \models \X \varphi &
\textit{iff} &
\mdp,\vscheduler,\vec{r}_1 \models \varphi\\
\mdp,\vscheduler,\xpath \models \varphi_1 \U \varphi_2 &
\textit{iff} &
\exists j \geq 0. \Big(\mdp,\vscheduler,\vec{r}_j\models\varphi_2 \wedge \forall i \in [0, j). \
\mdp,\vscheduler,\vec{r}_i \models \varphi_1\Big)\\
\mdp,\vscheduler,\xpath \models \varphi_1 \U^{[k_1,k_2]} \, \varphi_2 &
\textit{iff} &
\exists j \in [k_1, k_2]. \Big(\mdp,\vscheduler,\vec{r}_j\models\varphi_2 \wedge \forall i 
\in [0, j). \mdp,\vscheduler,\vec{r}_i \models \varphi_1\Big)
\end{array}
\]
}

\noindent where $\xpath=\vec{r}_0 \vec{r}_1\cdots$ with 
$\vec{r}_i=((\mode_{i,1},\state_{i,1}),\ldots,(\mode_{i,n},\state_{i,n}))$ is a 
path of $\mdp^{\vscheduler}$; formulas $\varphi$, $\varphi_1$, 
and $\varphi_2$ are \NHyperPCTL state formulas, and $k_1 
\leq k_2\in\naturalszero$.

\iflong
\subsection{The Expressiveness Power of HyperPCTL}

For MDPs with $|\Act(\state)|=1$ 
for each of its states $\state$, the \HyperPCTL semantics reduces to the one proposed in~\cite{ab18} for DTMCs.
%

For MDPs with non-determinism, the standard \PCTL semantics defines that in order to satisfy a \PCTL
formula $\pr_{\sim c} (\psi)$ in a given MDP state $\state$, {\em all}
schedulers should induce a DTMC that satisfies $\pr_{\sim c} (\psi)$
in $\state$. Though it should hold for \emph{all} schedulers, it is known
that there exist minimal and maximal schedulers that are
non-probabilistic and memoryless, therefore it is sufficient to
restrict the reasoning to such schedulers. Since for MDPs with
finite state and action spaces, the number of such schedulers is
finite, \PCTL model checking for MDPs is decidable. Given this analogy, one 
would expect that \HyperPCTL model checking should be decidable, but it is not.

\newcommand{\thrmundec}{\NHyperPCTL model checking for MDPs is in general 
undecidable.}

\begin{theorem}
\label{thrm:undec}
\thrmundec 
\end{theorem}

What is the source of increased expressiveness that makes \NHyperPCTL 
undecidable? State quantification cannot be the source, as the state space is 
finite and thus there are finitely many possible state quantifier 
instantiations.

Assume an MDP $\mdp= (\states, \Act, \P, \AP, L)$ with a state
$\state\in\states$ that is uniquely labelled by the proposition
$\textit{init}\in L(\state)$, and let $a,b\in\AP$. In PCTL, each
probability bound needs to be satisfied under all schedulers. For
example:
  \[
  \mdp,\state \models_{\PCTL} \pr_{< 0.5}(a\U b) 
  \textit{\ iff \ }
  \mdp\models_{\NHyperPCTL}\forall\hscheduler.\forall\hstateof{}{\hscheduler}.(\propof{\textit{init}}{\hstate}\rightarrow \pr(a\U b)< 0.5)
  \]
  Alternatively, we can state:
{\small
\[
  \mdp,\state \models_{\PCTL} \pr_{{<} 0.5}(a\U b) 
  \ \ \textit{iff} \ \
  \forall \scheduler{\in}\schedulers{\mdp}\!\!.
  \mdp,(\scheduler),((\start_\scheduler(\state),\state))\models_{\NHyperPCTL} \pr(a\mathcal{U} b){<} 0.5
\]
}

\noindent where $\start_\scheduler(\state)$ is the starting mode of scheduler 
$\scheduler$ in state $\state$.
Generally, the \NHyperPCTL fragment which starts with a single 
universal scheduler quantifier and contains a single bound on a single 
probability operator is still decidable.
  However, when a \PCTL formula has several probability bounds, its satisfaction 
requires each bound to be satisfied by all schedulers \emph{independently}. For 
example,
  \begin{eqnarray*}
  &&\mdp,\state \models_{\PCTL} \Big(\pr_{< 0.5}(a\U b) \; \vee \;\pr_{> 
0.5}(a\U 
b)\Big)
  \quad \textit{iff}\\
  &&\forall \scheduler\in\schedulers{\mdp}. \quad
  \mdp,(\scheduler),((\start_\scheduler(\state),\state))\models_{\NHyperPCTL} \pr(a\U b)< 0.5
  \quad \textit{or}\\
  &&\phantom{\forall \scheduler\in\schedulers{\mdp}.}\quad \mdp,(\scheduler),((\start_\scheduler(\state),\state))\models_{\NHyperPCTL} \pr(a\U b)> 0.5
  \end{eqnarray*}
This is \emph{not} equivalent to the $\NHyperPCTL$ formula 
  \[
  \mdp\models_{\NHyperPCTL}\forall\hscheduler.\forall\hstateof{}{\hscheduler}.(\propof{\textit{init}}{\hstate}\rightarrow(\pr(\propof{a}{\hstate}\U \propof{b}{\hstate})< 0.5 \vee \pr(\propof{a}{\hstate}\U \propof{b}{\hstate})> 0.5))
  \]
  which states that the probability is either less than or larger than $0.5$ under all schedulers, which is true if \emph{there exists no scheduler} under which the probability is $0.5$ (see also \cite{baier_acta12}). Thus, even for a fragment restricted to universal scheduler quantification, combinations of probability bounds allows \NHyperPCTL to express existential \emph{scheduler synthesis} problems.


Finally, consider a scheduler quantifier followed by state quantifiers, 
whose scope may contain probability expressions. This means we start several 
``experiments'' in parallel, each one represented by a state quantifier. 
However, we may use in all experiments the \emph{same scheduler}. Informally, 
this allows us to express the existence or absence of schedulers with certain 
probabilistic hyperproperties for the induced DTMCs. It would however also make 
sense to flip this quantifier order, such that state quantifiers are followed 
by scheduler quantifiers. This would mean, that we can use different schedulers 
in the different concurrently running experiments. This would be meaningful 
e.g. when users can provide input to the system, i.e. when the scheduler choice 
lies by the ``observers'' of the individual experiments, and they can adapt 
their schedulers to observations made in the other concurrently running 
experiments.

%

\else

For MDPs with $|\Act(\state)|=1$ 
for each of its states $\state$, the \HyperPCTL semantics reduces to the one proposed in~\cite{ab18} for DTMCs.

For MDPs with non-determinism, the standard \PCTL semantics defines that in order to satisfy a \PCTL
formula $\pr_{\sim c} (\psi)$ in a given MDP state $\state$, {\em all}
schedulers should induce a DTMC that satisfies $\pr_{\sim c} (\psi)$
in $\state$. Though it should hold for \emph{all} schedulers, it is known
that there exist minimal and maximal schedulers that are
non-probabilistic and memoryless, therefore it is sufficient to
restrict the reasoning to such schedulers. Since for MDPs with
finite state and action spaces, the number of such schedulers is
finite, \PCTL model checking for MDPs is decidable. Given this analogy, one 
would expect that \HyperPCTL model checking should be decidable, but it is not.

\newcommand{\thrmundec}{\NHyperPCTL model checking for MDPs is in general 
undecidable.}

\begin{theorem}
\label{thrm:undec}
\thrmundec 
\end{theorem}

\ \\

\fi

\section{Applications of HyperPCTL on MDPs}
\label{sec:app}

\begin{wrapfigure}[12]{r}{4.9cm}
\vspace{-1cm}
\begin{lstlisting}[style=CStyle,tabsize=2,language=ML,basicstyle=\scriptsize
%    ,escapechar=/
  ]
void mexp(){
  c = 0; d = 1; i = k;
  while (i >= 0){
    i = i-1; c = c*2;
    d = (d*d) % n;
    if (b(i) = 1)
      c = c+1;
      d = (d*a) % n;
    }
}
/************/
t = new Thread(mexp()); 
j = 0; m = 2 * k;
while (j < m & !t.stop) j++;
/************/
\end{lstlisting}
\vspace{-.4cm}
\caption{Modular exponentiation.}
\label{fig:modexp}
\end{wrapfigure}

\paragraph{Side-channel timing leaks} open a channel to an attacker to 
infer the value of a secret by observing the execution time of a
function. For example, the heart of the RSA public-key encryption
algorithm is the modular exponentiation algorithm that computes $(a^b
\mod n)$, where $a$ is an integer representing the plaintext and $b$
is the integer encryption key.  A careless implementation can leak $b$
through a probabilistic scheduling channel (see
Fig.~\ref{fig:modexp}). This program is not secure since the two
branches of the \emph{if} have different timing behaviors. Under a
fair execution scheduler for parallel threads, an attacker thread can
infer the value of $b$ by running in parallel to a modular
exponentiation thread and iteratively incrementing a counter variable
until the other thread terminates (lines 12-14).
To model this program by an MDP, we can use two nondeterministic actions for the two branches of the \emph{if} statement, such that the choice of different schedulers corresponds to 
the choice of different bit configurations \texttt{b(i)} for the key \texttt{b}. 
This algorithm should satisfy the following property:
the probability of observing a concrete value in the counter \texttt{j} should 
be independent of the bit configuration of the secret key \texttt{b}:  
%
$$
\forall \hscheduler_1. \forall \hscheduler_2. \forall 
\hstate(\hscheduler_1).\forall \hstate'(\hscheduler_2). \Big(init_{\hstate} 
\, \wedge \, init_{\hstate'} \Big) \; 
\Rightarrow \;  \bigwedge_{l=0}^{m}\Big(\pr(\F (j=l)_{\hstate} ) = \pr(\F (j=l)_{\hstate'} )\Big)
$$
%
%

\begin{wrapfigure}[8]{r}{4.9cm}
\vspace{-1cm}
\begin{lstlisting}[style=CStyle,tabsize=2,language=ML,basicstyle=\scriptsize
%    ,escapechar=/
  ]
int str_cmp(char * r){
  char * s = 'Bg\$4\0';
  i = 0;
  while (s[i] != '\0'){
    i++;
    if (s[i]!=r[i]) return 0;
    }
    return 1;
}
\end{lstlisting}
\vspace{-.5cm}
\caption{String comparison.}
\label{fig:string}
\end{wrapfigure}

Another example of timing attacks that can be implemented through a 
probabilistic scheduling side channel is password verification which is 
typically implemented by comparing an input string with another confidential 
string (see Fig~\ref{fig:string}). Also here, an attacker thread can measure the 
time necessary to break the loop, and use this information to infer the prefix of 
the input string matching the secret string.

\paragraph{Scheduler-specific observational determinism policy} 
(SSODP)~\cite{nsh13} is a confidentiality policy in multi-threaded
programs that defends against an attacker that chooses an appropriate
scheduler to control the set of possible traces. In particular, given
any scheduler and two initial states that are indistinguishable with
respect to a secret input (i.e., low-equivalent), any two executions
from these two states should terminate in low-equivalent states with equal
probability. Formally, given a proposition $h$ representing a secret:
%
%
\begin{align*}
\forall \hscheduler.\forall \hstate(\hscheduler).\forall \hstate'(\hscheduler). 
\big(h_{\hstate} \oplus h_{\hstate'} \big) \, \Rightarrow \, 
\bigwedge_{l \in L}\big( & \pr (\F \propof{l}{\hstate}) = \pr (\F \propof{l}{\hstate'})\big)
\end{align*}
where $l\in L$ are atomic propositions that classify low-equivalent states and 
$\oplus$ is the exclusive-or operator. A stronger variation of this policy is 
that the executions are stepwise low-equivalent:
\begin{align*}
\forall \hscheduler.\forall \hstate(\hscheduler).\forall \hstate'(\hscheduler). 
\big(h_{\hstate} \oplus h_{\hstate'} \big) \, \Rightarrow \,
\pr\G \big(\bigwedge_{l \in L}\big((\pr\X \propof{l}{\hstate}) = (\pr\X 
\propof{l}{\hstate'}) \big)\big) = 1.
\end{align*}
\paragraph{Probabilistic conformance} describes how well a model and 
an implementation conform with each other with respect to a specification.
As an example, consider a 6-sided die. The probability to obtain one possible 
side of the die is $1/6$. We would like to synthesize a protocol that simulates 
the 6-sided die behavior only by repeatedly tossing a fair coin. We know that 
such an implementation exists~\cite{KY76}, but our aim is to find 
such a solution automatically by modeling the die as a DTMC and by using an MDP 
to model all the possible coin-implementations with a given maximum number of 
states, including 6 absorbing final states to model the outcomes. In the MDP, we associate to each state a set of possible nondeterministic actions, each of them choosing two states as successors with equal probability $1/2$.
Then, each scheduler corresponds to a particular
implementation. Our goal is to check whether there exists a scheduler that 
induces a DTMC over the MDP, such that repeatedly tossing a coin simulates die-rolling with equal probabilities for the different outcomes:
$$
\exists \hscheduler.  \forall 
\hstate(\hscheduler). \exists \hstate'(\hscheduler). \Big(init_{\hstate} 
\, \wedge \, init_{\hstate'} \Big) \; 
\Rightarrow \;  \bigwedge_{l=1}^{6}\Big(\pr(\F (die=l)_{\hstate} ) = \pr(\F (die=l)_{\hstate'} )\Big)
$$
\section{HyperPCTL Model Checking for Non-probabilistic Memoryless Schedulers}
\label{sec:memless}

Due to the undecidability of model checking \NHyperPCTL formulas for MDPs, we noe restrict ourselves the semantics, where scheduler
quantification ranges over non-probabilistic memoryless schedulers
only. It is easy to see that this restriction makes the model checking
problem decidable, as there are only finitely many such schedulers that can be enumerated. Regarding 
complexity, we have the following property.

\newcommand{\thrmnpc}{The problem to decide for MDPs the truth of \NHyperPCTL formulas with a single
existential (respectively, universal) scheduler quantifier over non-probabilistic 
memoryless schedulers is \comp{NP-complete} (respectively, 
\comp{coNP-complete}) in the state set size of the given MDP.
}
\begin{theorem}
\label{thrm:npc}
\thrmnpc
\end{theorem}

Next we propose an 
SMT-based technique for solving the model checking problem for non-probabilistic memoryless scheduler domains, and for the simplified case of having a single scheduler quantifier; the general case for an arbitrary number of scheduler quantifiers is similar, but a bit more involved, so the simplified setting might be more suitable for understanding the basic ideas.

\SetCustomAlgoRuledWidth{9.5cm}
\begin{wrapfigure}[17]{r}{7.25cm}
  \scalebox{0.75}{
    \begin{minipage}{14cm}
  \begin{algorithm}[H]
  \caption{Main SMT encoding algorithm}
  \label{algo:smt_main}
  \Input{$\mdp=(\states, \Act,\P, \AP, L)$: MDP;\\$Q\hscheduler. Q_1\hstateof{1}{\hscheduler}.\ldots Q_n\hstateof{n}{\hscheduler}. \nqform$:~\NHyperPCTL~formula.}
  \Output{Whether $\mdp$ satisfies the input formula.}
  \SetKwFunction{Main}{Main}
  \Fn{Main{$(\mdp$,\ \ $Q\hscheduler.\ Q_1\hstateof{1}{\hscheduler}. \ldots Q_n\hstateof{n}{\hscheduler}.\ \nqform)$}}{
    $E$ := $\bigwedge_{\state\in\states}(\bigvee_{\action\in\Act(\state)}\scheduler_{\state}=\action)$ \tcp{scheduler choice}\label{line:scheduler}
    \If{$Q$ is existential\label{line:exists}}{
      $E$ := $E\wedge$ Semantics$(\mdp$, $\nqform$, $n)$\;\label{line:existsem}
      $E$ := $E\wedge$ Truth$(\mdp$, $\exists\hscheduler.\ Q_1\hstateof{1}{\hscheduler}. \ldots Q_n\hstateof{n}{\hscheduler}.\ \nqform)$\;\label{line:existtrue}
      \lIf{check$(E)$ = SAT\label{line:existsatA}}{\Return{TRUE}}\lElse{\Return{FALSE}\label{line:existsatB}}
    }\ElseIf{$Q$ is universal\label{line:forallA}}{
      \tcp{$\overline{Q}_i$ is $\forall$ if $Q_i=\exists$ and $\exists$ else}
      $E$ := $E\wedge$ Semantics$(\mdp,\neg\nqform,n)$\;
      $E$ := $E\wedge$ Truth$(\mdp$, $\exists\hscheduler. \overline{Q}_1\hstateof{1}{\hscheduler}. \ldots \overline{Q}_n\hstateof{n}{\hscheduler}. \neg\nqform)$\;
      \lIf{check$(E)$ = SAT}{\Return{FALSE}}\lElse{\Return{TRUE}\label{line:forallB}}
    }
  }
  \end{algorithm}
  \end{minipage}
}
\end{wrapfigure}

The main method listed in Algorithm
\ref{algo:smt_main} constructs a formula $E$ that is satisfiable if and only if the input MDP $\mdp$ satisfies the input \NHyperPCTL formula with a single scheduler quantifier over the non-probabilistic memoryless scheduler domain. Let us first deal with the case that the scheduler quantifier is \emph{existential}.
In line \ref{line:scheduler} we encode possible instantiations $\scheduler$ for the scheduler variable $\hscheduler$, for which we use a variable $\scheduler_s$ for each MDP state $\state\in\states$ to encode which action is chosen in that state.
In line \ref{line:existsem} we encode the meaning of the quantifier-free inner part $\nqform$ of the input formula, whereas line \ref{line:existtrue} encodes the meaning of the state quantifiers, i.e. for which sets of composed states $\nqform$ needs to hold in order to satisfy the input formula. In lines \ref{line:existsatA}--\ref{line:existsatB} we check the satisfiability of the encoding and return the corresponding answer.
Formulas with a \emph{universal} scheduler quantifier $\forall\hscheduler.\formula$ are semantically equivalent to $\neg\exists\hscheduler.\neg\formula$. We make use of this fact in lines \ref{line:forallA}--\ref{line:forallB} to check first the satisfaction of an encoding for $\exists\hscheduler.\neg\formula$ and return the inverted answer.

\SetCustomAlgoRuledWidth{1.04\linewidth}
\begin{figure}[t]
  \scalebox{0.9}{
    \begin{minipage}{1.1\linewidth}
\begin{algorithm}[H]
  \caption{SMT encoding for the meaning of the input formula}
  \label{algo:smt_sub}
  \Input{$\mdp=(\states, \Act,\P, \AP, L)$: MDP;\\
    $\formula$: quantifier-free \NHyperPCTL formula or expression;\\
    $n$: number of state variables in $\formula$.}
  \Output{SMT encoding of the meaning of $\formula$ in the $n$-ary self-composition of $\mdp$.}
  \SetKwFunction{Semantics}{Semantics}
  \Fn{Semantics{$(\mdp$, $\formula$, $n)$}}{
    \lIf{$\formula$ is $\tru$\label{line:semtrue}}{
      $E$ := $\bigwedge_{\vstate\in\states^n}\btruth{\vstate}{\formula}$%
    }
    \ElseIf{$\formula$ is $\propof{a}{\hstate_i}$\label{line:semproposition}}{
      $E$ := $(\bigwedge_{\vstate\in\states^n,\ a\in L(\state_i) }(\btruth{\vstate}{\formula})) \wedge (\bigwedge_{\vstate\in\states^n,\ a\notin L(\state_i) }(\neg \btruth{\vstate}{\formula}))$;%
    }
    \ElseIf{$\formula$ is $\formula_1\wedge\formula_2$\label{line:semconjunction}}{
      $E$ := Semantics$(\mdp,\formula_1,n) \wedge$ Semantics$(\mdp, \formula_2,n)\wedge$\\
      \ $\bigwedge_{\vstate\in\states^n}((\btruth{\vstate}{\formula}{\wedge} \btruth{\vstate}{\formula_1}{\wedge} \btruth{\vstate}{\formula_2}) \vee (\neg \btruth{\vstate}{\formula} {\wedge} (\neg \btruth{\vstate}{\formula_1}{\vee} \neg \btruth{\vstate}{\formula_2})))$\;
    }
    \ElseIf{$\formula$ is $\neg\formula'$\label{line:semnegation}}{
      $E$ := Semantics$(\mdp,\formula',n)\wedge \bigwedge_{\vstate\in\states^n}(\btruth{\vstate}{\formula} \oplus \btruth{\vstate}{\formula'})$\;
    }
    \ElseIf{$\formula$ is $\formula_1 < \formula_2$\label{line:semcomparison}}{
      $E$ := Semantics$(\mdp$, $\formula_1,n)\wedge$ Semantics$(\mdp$, $\formula_2,n)\wedge$\\
      \quad$\bigwedge_{\vstate\in\states^n}((\btruth{\vstate}{\formula}\wedge \ptruth{\vstate}{\formula_1} < \ptruth{\vstate}{\formula_2})\vee(\neg \btruth{\vstate}{\formula}\wedge \ptruth{\vstate}{\formula_1} \geq \ptruth{\vstate}{\formula_2}))$\;
    }
    \ElseIf{$\formula$ is $\pr(\Next\formula')$\label{line:semnext}}{
      $E$ := Semantics$(\mdp$, $\formula',n)\wedge$\label{line:semnextB}\\
      \quad$\bigwedge_{\vstate\in\states^n}\big((\btoi{\vstate}{\formula'}=1\wedge\btruth{\vstate}{\formula'})\vee (\btoi{\vstate}{\formula'}=0\wedge\neg\btruth{\vstate}{\formula'})\big)$\label{line:semnextC}\;
      \ForEach{$\vstate=(\state_1,\ldots,\state_n)\in\states^n$\label{line:statechoice}}{
      \ForEach{$\vec{\action}=(\action_1,\ldots,\action_n)\in\Act(\state_1)\times\ldots\times\Act(\state_n)$\label{line:schedulerchoice}}{
          $E$ := $ E \wedge \big( \big[ \bigwedge_{i=1}^{n} \scheduler_{\state_i}=\action_i \big] \rightarrow \big[ \ptruth{\vstate}{\formula} = $\\
          $\sum_{\vstate'\in\supp(\alpha_1)\times\ldots\times\supp(\alpha_n)} ((\Pi_{i=1}^n \P(\state_i,\alpha_i,\state_i'))\cdot \btoi{\vstate'}{\formula'})\big]\big)$;\label{line:nextprob}
      }
      }
    }
    \lElseIf{$\formula$ is $\pr(\formula_1\U \formula_2)$\label{line:semunboundeduntil}}{
      $E$ := SemanticsUnboundedUntil($\mdp$, $\formula$, $n$)%
    }
    \lElseIf{$\formula$ is $\pr(\formula_1\U^{[k_1,k_2]}\formula_2)$\label{line:semboundeduntil}}{
      $E$ := SemanticsBoundedUntil($\mdp$, $\formula$, $n$)%
    }
    \lElseIf{$\formula$ is $c$\label{line:semconstant}}{
      $E$ := $\bigwedge_{\vstate\in\states^n}(\ptruth{\vstate}{\formula}=c)$%
    }
    \ElseIf{$\formula$ is $\formula_1\textit{ op }\formula_2$ \quad \texttt{/*} $\textit{op}\in\{+,-,*\}$ \texttt{*/} \quad \label{line:semarithmetic}}{
      $E$ := Semantics$(\mdp$, $\formula_1,n)\wedge$ Semantics$(\mdp$, $\formula_2,n)\wedge$
      $\bigwedge_{\vstate\in\states^n}(\ptruth{\vstate}{\formula}=(\ptruth{\vstate}{\formula_1}\textit{ op } \ptruth{\vstate}{\formula_2}))$;
    }
    \Return{$E$};
  }
\end{algorithm}
  \end{minipage}
}
\end{figure}

The Semantics method, shown in Algorithm \ref{algo:smt_sub}, applies structural recursion to encode the meaning of its quantifier-free input formula. As variables, the encoding uses (1) propositions \erika{$\ptruth{\vstate}{\nqform}\in\{0,1\}$}{$\btruth{\vstate}{\nqform}\in\{\tru,\fals\}$} to encode the truth of each Boolean sub-formula $\nqform$ of the input formula in each state $\vstate\in\states^n$ of the $n$-ary self-composition of $\mdp$, (2) numeric variables $\ptruth{\vstate}{\prform}\in[0,1]\subseteq\mathbb{R}$ to encode the value of each probability expression $\prform$ in the input formula in the context of each composed state $\vstate\in\states^n$, \erika{}{ (3) variables $\btoi{\vstate}{\prform}\in\{0,1\}$ to encode truth values in a pseudo-Boolean form, i.e. we set $\btoi{\vstate}{\prform}=1$ for $\btruth{\vstate}{\nqform}=\tru$ and $\ptruth{\vstate}{\prform}=0$ else and (4) variables $d_{\vstate,\formula}$ to encode the existence of a loop-free path from state $\vstate$ to a state satisfying $\formula$.}

There are two base cases: the Boolean constant \texttt{true} holds in all states (line \ref{line:semtrue}), whereas atomic propositions hold in exactly those states that are labelled by them (line \ref{line:semproposition}). 
For conjunction (line \ref{line:semconjunction}) we recursively encode the truth values of the operands and state that the conjunction is true iff both operands are true. For negation (line \ref{line:semnegation}) we again encode the meaning of the operand recursively and flip its truth value. For the comparison of two probability expressions (line \ref{line:semcomparison}), we recursively encode the probability values of the operands and state the respective relation between them for the satisfaction of the comparison.

The remaining cases encode the semantics of probability expressions. The cases 
for constants (line \ref{line:semconstant}) and arithmetic operations (line 
\ref{line:semarithmetic}) are straightforward. For the probability 
$\pr(\Next\formula')$ (line \ref{line:semnext}), we encode the Boolean value of 
$\formula'$ in the variables $\btruth{\vstate}{\formula'}$ (line 
\ref{line:semnextB}), turn them into \erika{arithmetic}{pseudo-Boolean} values 
$\btoi{\vstate}{\formula'}$ ($1$ for true and $0$ for false, line 
\ref{line:semnextC}), and state that for each composed state, the probability 
value of $\pr(\Next\formula')$ is the sum of the probabilities to get to a successor state where the operand $\formula'$ holds; 
since the successors and their probabilities are scheduler-dependent, we need to 
iterate over all scheduler choices and use $\supp(\alpha_i)$ to denote the support $\{\state\in\states\,\ \,\alpha_i(\state)>0\}$ of the distribution $\alpha_i$ (line \ref{line:schedulerchoice}).
The encodings for the probabilities of unbounded until formulas (line \ref{line:semunboundeduntil}) and bounded until formulas (line \ref{line:semboundeduntil}) are listed in Algorithm \ref{algo:smt_sub_unbounded_until}  and \ref{algo:smt_sub_bounded_until}, respectively.

\SetCustomAlgoRuledWidth{1.04\linewidth}
\begin{figure}[t]
  \scalebox{0.9}{
    \begin{minipage}{1.1\linewidth}
\begin{algorithm}[H]
  \caption{SMT encoding for the meaning of unbounded until formulas}
  \label{algo:smt_sub_unbounded_until}
  \Input{$\mdp=(\states, \Act,\P, \AP, L)$: MDP;\quad $\formula$: \NHyperPCTL unbounded until formula of the form $\pr(\formula_1\U \formula_2)$;\quad $n$: number of state variables in $\formula$.}
  \Output{SMT encoding of $\formula$'s meaning in the $n$-ary self-composition of $\mdp$.}
  \SetKwFunction{SemanticsUnboundedUntil}{SemanticsUnboundedUntil}
  \Fn{SemanticsUnboundedUntil{$(\mdp$, $\formula=\pr(\formula_1\U\formula_2)$, $n)$}}{
      $E$ := Semantics$(\mdp$, $\formula_1,n)\wedge$ Semantics$(\mdp$, $\formula_2,n)$\label{line:unboundedA}\;
      \ForEach{$\vstate=(\state_1,\ldots,\state_n)\in\states^n$\label{line:unboundedD}}{
        $E$ := $E\wedge (\btruth{\vstate}{\formula_2} \rightarrow \ptruth{\vstate}{\formula}{=}1)\wedge ((\neg \btruth{\vstate}{\formula_1}\wedge \neg \btruth{\vstate}{\formula_2}) \rightarrow \ptruth{\vstate}{\formula}{=}0)$\label{line:unboundedE}\;
        \ForEach{$\vec{\action}=(\action_1,\ldots,\action_n)\in\Act(\state_1)\times\ldots\times\Act(\state_n)$\label{line:unboundedF}}{
          $E$ := $ E \wedge \Big( \big[ \btruth{\vstate}{\formula_1} \wedge \neg \btruth{\vstate}{\formula_2} \wedge \bigwedge_{i=1}^{n} \scheduler_{\state_i}=\action_i \big] \rightarrow $\label{line:unboundedG}\\
          $\big[ \ptruth{\vstate}{\formula} = \sum_{\vstate'\in\supp(\alpha_1)\times\ldots\times\supp(\alpha_n)} ((\Pi_{i=1}^n \P(\state_i,\alpha_i,\state_i'))\cdot \ptruth{\vstate'}{\formula})\wedge$\label{line:unboundedH}\\
          $(\ptruth{\vstate}{\formula}{>}0\rightarrow (\bigvee_{\vstate'\in\supp(\alpha_1)\times\ldots\times\supp(\alpha_n)} (\btruth{\vstate'}{\formula_2}{\vee} d_{\vstate,\formula_2}{>}d_{\vstate',\formula_2}))) \big] \Big) $\label{line:unboundedI}\;
        }
      }
    \Return{$E$};
  }
\end{algorithm}
  \end{minipage}
}
\end{figure}
For the probabilities $\pr(\formula_1\U\formula_2)$ to satisfy an unbounded until formula, the method SemanticsUnboundedUntil shown in Algorithm \ref{algo:smt_sub_unbounded_until} first encodes the meaning of the until operands (line \ref{line:unboundedA}). For each composed state $\vstate\in\states^n$, the probability of satisfying the until formula in $\vstate$ is encoded in the variable $\ptruth{\vstate}{\pr(\formula_1\U\formula_2)}$. If the second until-operand $\formula_2$ holds in $\vstate$ then this probability is $1$ and if none of the operands are true in $\vstate$ then it is $0$ (line \ref{line:unboundedE}). Otherwise, depending on the scheduler $\scheduler$ of $\mdp$ (line \ref{line:unboundedF}), the value of $\ptruth{\vstate}{\pr(\formula_1\U\formula_2)}$ is a sum, adding up for each successor state $\vstate'$ of $\vstate$ the probability to get from $\vstate$ to $\vstate'$ in one step times the probability to satisfy the until-formula on paths starting in $\vstate'$ (line \ref{line:unboundedH}). However, these encodings work only when at least one state satisfying $\formula_2$ is reachable from $\vstate$ with a positive probability: for any bottom SCC whose states all violate $\formula_2$, the probability $\pr(\formula_1\U\formula_2)$ is obviously $0$, however, assigning any fixed value from $[0,1]$ to all states of this bottom SCC would yield a fixed-point for the underlying equation system. To assure correctness, in line \ref{line:unboundedI} we enforce smallest fixed-points by requiring that if $\ptruth{\vstate}{\pr(\formula_1\U\formula_2)}$ is positive then there exists a loop-free path from $\vstate$ to any state satisfying $\formula_2$. In the encoding of this property we use fresh variables $d_{\vstate,\formula_2}$ and require a path over states with strong monotonically decreasing $d_{\vstate,\formula_2}$-values to a $\formula_2$-state (where the decreasing property serves to exclude loops). The domain of the distance-variables $d_{\vstate,\formula_2}$ can be e.g. integers, rationals or reals; the only restriction is that is should contain at least $|\states|^n$ ordered values. Especially, it does not need to be lower bounded (note that each solution assigns to each $d_{\vstate,\formula_2}$ a fixed value, leading a finite number of distance values).

\SetCustomAlgoRuledWidth{1.04\linewidth}
\begin{figure}[t]
  \scalebox{0.9}{
    \begin{minipage}{1.1\linewidth}
\begin{algorithm}[H]
  \caption{SMT encoding for the meaning of bounded until formulas}
  \label{algo:smt_sub_bounded_until}
  \Input{$\mdp=(\states, \Act,\P, \AP, L)$: MDP;\quad $\formula$: \NHyperPCTL bounded until formula of the form $\pr(\formula_1\U^{[k_1,k_2]}\formula_2)$;\quad $n$: number of state variables in $\formula$.}
  \Output{SMT encoding of $\formula$'s meaning in the $n$-ary self-composition of $\mdp$.}
  \SetKwFunction{SemanticsBoundedUntil}{SemanticsBoundedUntil}
  \Fn{SemanticsBoundedUntil{$(\mdp$, $\formula=\pr(\formula_1\U^{[k_1,k_2]}\formula_2)$, $n)$}}{
    \If{$k_2=0$\label{line:boundedA}}{
      $E$ := Semantics$(\mdp,\formula_1,n)\wedge$ Semantics$(\mdp,\formula_2,n)$\label{line:boundedB}\;
      \ForEach{$\vstate=(\state_1,\ldots,\state_n)\in\states^n$\label{line:boundedC}}{
        $E$ := $E\wedge (\btruth{\vstate}{\formula_2} {\rightarrow} \ptruth{\vstate}{\formula}{=}1) \wedge (\neg \btruth{\vstate}{\formula_2} {\rightarrow} \ptruth{\vstate}{\formula}{=}0)$\label{line:boundedD}\;
      }
    }
    \ElseIf{$k_1=0$\label{line:boundedE}}{
      $E$ := SemanticsBoundedUntil$(\mdp$, $\pr(\formula_1\U^{[0,k_2-1]}\formula_2),n)$\label{line:boundedF}\;
      \ForEach{$\vstate=(\state_1,\ldots,\state_n)\in\states^n$\label{line:boundedG}}{
        $E$ := $E\wedge (\btruth{\vstate}{\formula_2} {\rightarrow} \ptruth{\vstate}{\formula}{=}1)\wedge ((\neg \btruth{\vstate}{\formula_1}\wedge \neg \btruth{\vstate}{\formula_2}) {\rightarrow} \ptruth{\vstate}{\formula}{=}0)$\label{line:boundedH}\;
        \ForEach{$\vec{\action}=(\action_1,\ldots,\action_n)\in\Act(\state_1)\times\ldots\times\Act(\state_n)$\label{line:boundedI}}{
          $E$ := $ E \wedge \Big( \big[ \ \ \btruth{\vstate}{\formula_1} \wedge \neg \btruth{\vstate}{\formula_2} \wedge \bigwedge_{i=1}^{n} \scheduler_{\state_i}=\action_i \ \ \big] \rightarrow  \big[ \ptruth{\vstate}{\formula} = $\label{line:boundedJ}\\
          $\sum_{\vstate'\in\supp(\alpha_1)\times\ldots\times\supp(\alpha_n)} ((\Pi_{i=1}^n \P(\state_i,\alpha_i,\state_i'))\cdot \ptruth{\vstate'}{\pr(\formula_1\U^{[0,k_2-1]}\formula_2)}) \ \ \big] \Big) $\label{line:boundedK}\;
        }
      }
    }
    \ElseIf{$k_1>0$\label{line:boundedL}}{
      $E$ := SemanticsBoundedUntil$(\mdp$, $\pr(\formula_1\U^{[k_1-1,k_2-1]}\formula_2),n)$\label{line:boundedM}\;
      \ForEach{$\vstate=(\state_1,\ldots,\state_n)\in\states^n$\label{line:boundedN}}{
        $E$ := $E\wedge (\neg \btruth{\vstate}{\formula_1} \rightarrow \ptruth{\vstate}{\formula}=0)$\label{line:boundedO}\;
        \ForEach{$\vec{\action}=(\action_1,\ldots,\action_n)\in\Act(\state_1)\times\ldots\times\Act(\state_n)$\label{line:boundedP}}{
          $E$ := $ E \wedge \Big( \big[\ \  \btruth{\vstate}{\formula_1} \wedge \bigwedge_{i=1}^{n} \scheduler_{\state_i}=\action_i \ \ \big] \rightarrow \big[ \ \ \ptruth{\vstate}{\formula} = $\label{line:boundedQ}\\
          $\sum_{\vstate'\in\supp(\alpha_1)\times\ldots\times\supp(\alpha_n)} ((\Pi_{i=1}^n \P(\state_i,\alpha_i,\state_i'))\cdot \ptruth{\vstate'}{\pr(\formula_1\U^{[k_1-1,k_2-1]}\formula_2)}) \ \ \big] \Big) $\label{line:boundedR}\;
        }
      }
    }
    \Return{$E$};
  }
\end{algorithm}
  \end{minipage}
}
\end{figure}

\SetCustomAlgoRuledWidth{9.3cm}
\begin{wrapfigure}[10]{r}{8.2cm}
\iflong
\else
\vspace*{-4ex}
\fi
  \scalebox{0.8}{
    \begin{minipage}{10cm}
\begin{algorithm}[H]
  \caption{SMT encoding of the truth of the input formula}
  \label{algo:smt_truth}
  \Input{$\mdp=(\states, \Act,\P, \AP, L)$: MDP;\\  $\exists\hscheduler. Q_1\hstateof{1}{\hscheduler}. \ldots Q_n\hstateof{n}{\hscheduler}. \nqform$: \NHyperPCTL formula.}
  \Output{Encoding of the truth of the input formula in $\mdp$.}
  \SetKwFunction{Truth}{Truth}
  \Fn{Truth{$(\mdp$, $\exists\hscheduler.\ Q_1\hstateof{1}{\hscheduler}.\ \ldots\ Q_n\hstateof{n}{\hscheduler}.\ \nqform)$}}{
    \ForEach{$i=1,{\ldots},n$\label{line:truthA}}{
      \lIf{$Q_i=\forall$}{$B_i:=$''$\bigwedge_{\state_i\in\states}$''\label{line:truthB}}
      \lElse{$B_i:=$''$\bigvee_{\state_i\in\states}$''\label{line:truthC}}
      }
    \Return{$B_1\ \ldots \ B_n\ \btruth{(\state_1,\ldots,\state_n)}{\nqform}$};
  }
\end{algorithm}
  \end{minipage}
}
\end{wrapfigure}
The SemanticsBoundedUntil method, listed in Algorithm \ref{algo:smt_sub_bounded_until}, encodes the probability $\pr(\formula_1\U^{[k_1,k_2]}\formula_2)$ of a bounded until formula in the numeric variables $\ptruth{\vstate}{\pr(\formula_1\U^{[k_1,k_2]}\formula_2)}$ for all (composed) states $\vstate\in\states^ n$ and recursively reduced
time bounds. There are three main cases: (i) the satisfaction of $\formula_1\U^{[0,k_2-1]}\formula_2$ requires to satisfy $\formula_2$ immediately (lines \ref{line:boundedA}--\ref{line:boundedD}); (ii) $\formula_1\U^{[0,k_2-1]}\formula_2$ can be satisfied by either satisfying $\formula_2$ immediately or satisfying it later, but in the latter case $\formula_1$ needs to hold currently (lines \ref{line:boundedE}--\ref{line:boundedK}); (iii) $\formula_1$ has to hold and $\formula_2$ needs to be satisfied some time later (lines \ref{line:boundedL}--\ref{line:boundedR}). To avoid the repeated encoding of the semantics of the operands, we do it only when we reach case (i) where recursion stops (line \ref{line:boundedB}). 
For the other cases, we recursively encode the probability to reach a $\formula_2$-state over $\formula_1$ states where the deadlines are reduced with one step (lines \ref{line:boundedF} resp. \ref{line:boundedM}) and use these to fix the values of the variables $\ptruth{\vstate}{\pr(\formula_1\U^{[k_1,k_2]}\formula_2)}$, similarly to the unbounded case but under additional consideration of time bounds.

Finally, the Truth method listed in Algorithm~\ref{algo:smt_truth} encodes the meaning of the state quantification: it states for each universal quantifier that instantiating it with any MDP state should satisfy the formula (conjunction over all states in line~\ref{line:truthB}), and for each existential state quantification that at least one state should lead to satisfaction (disjunction in line~\ref{line:truthC}).

\newcommand{\thrmsound}{Algorithm \ref{algo:smt_main} returns a formula that is true iff its input \HyperPCTL formula is satisfied by the input MDP.}

\begin{theorem}
\label{thrm:sound}
\thrmsound
\end{theorem}

We note that the satisfiability of the generated SMT encoding for a formula with 
an existential scheduler quantifier does not only prove the truth of the formula 
but provides also a scheduler as witness, encoded in the solution of the SMT 
encoding. Conversely, unsatisfiability of the SMT encoding for a formula with a 
universal scheduler quantifier provides a counterexample scheduler.

\newcommand{\TA}{{\bf TA}\xspace}
\newcommand{\TS}{{\bf TS}\xspace}
\newcommand{\PW}{{\bf PW}\xspace}
\newcommand{\PC}{{\bf PC}\xspace}

\section{Evaluation}
\label{sec:results}

We developed a prototypical implementation of our algorithm in python,
with the help of several libraries. There is an extensive use of
\comp{STORMPY}~\cite{stormpy,DehnertJK017}, which provides efficient solution 
to parsing, building, and storage of MDPs. We 
used the SMT-solver Z3~\cite{dmb08} to solve the logical encoding generated by 
Algorithm~\ref{algo:smt_main}. All of our experiments were run on a MacBook Pro 
laptop with a 2.3GHz i7 processor with 32GB of RAM. The results are 
presented in Table~\ref{tab:results}.

As the first case study, we model and analyze information leakage in the 
modular exponentiation algorithm (function \texttt{modexp} in 
Fig.~\ref{fig:modexp}); the corresponding results in Table~\ref{tab:results} are 
marked by \TA. We experimented with 1, 2, and 3 bits for the
encryption key (hence, $m \in \{2, 4, 6\}$). The specification checks
whether there is a timing channel for all possible schedulers, which
is the case for the implementation in \texttt{modexp}.

Our second case study is verification of password leakage thorough the string
comparison algorithm (function
\texttt{str\_cmp} in Fig~\ref{fig:string}).  Here, we also experimented with $m \in \{2, 4,
6\}$; results in Table~\ref{tab:results} are denoted by \PW.

In our third case study,
we assume two concurrent processes. The first
process decrements the value of a secret $h$ by $1$ as long as the
value is still positive, and after this it sets a low variable $l$ to
$1$. A second process just sets the value of the same low variable $l$ to
$2$. The two threads run in parallel; as long as none of them
terminated, a fair scheduler chooses for each CPU cycle the next
executing thread.  As discussed in Section~\ref{sec:intro}, this MDP
opens a probabilistic thread scheduling channel and leaks the value of
$h$. We denote this case study by \TS in Table~\ref{tab:results}, and
compare observations for executions with different secret values $h_1$ and 
$h_2$ (denoted as $h=(h_1,h_2)$ in the table). There is an interesting
relation between the execution times for \TA and \TS. For example, although the 
MDP for \TA with $m=4$ has 60 reachable states and the MDP for \TS comparing
executions for $h=(0,15)$ has 35 reachable states, verification of \TS takes 20 
times more than \TA. We
believe this is because the MDP of \TS is twice deeper
than the MDP of \TA, making the SMT constraints more complex.

Our last case study is on probabilistic conformance, denoted \PC. The input 
is a DTMC that encodes the behavior of a 6-sided die as well as a structure of 
actions having probability distributions with two successor states each; these 
transitions can be pruned using a scheduler to obtain a DTMC which simulates 
the die outcomes using a fair coin. Given a fixed state space, we experiment 
with different numbers of transitions.  In particular, we started from the implementation in~\cite{KY76}
and then we added all the possible nondeterministic transitions from the first state 
to all the other states (s=0), from the first and second states to all the others
(s=0,1), and from the first, second, and third states to all the others
(s=0,1,2). Each time we were able not only to satisfy the formula, but also to obtain 
 the witness corresponding to the scheduler satisfying the property. 

Regarding the running times listed in Table~\ref{tab:results}, we note
that our implementation is only prototypical and there are possibilities for 
numerous optimizations. Most importantly, for purely existentially or purely 
universally quantified formulas, we could define a more efficient encoding with 
much less variables. However, it is clear that the running times for even 
relatively small MPDs are 
large. This is simply because of the high complexity of the verification of 
hyperproperties. In addition, the \HyperPCTL formulas in our case studies have 
multiple scheduler and/or state quantifiers, making the problem significantly 
more difficult. 

\begin{table}[t]
\centering
  \scalebox{0.83}{
    
\begin{tabular}{|c|c||r|r|r||@{\,}c@{\,}|@{\,}c@{\,}|@{\,}c@{\,}|@{\,}c@{\,}|}
\hline
\multicolumn{2}{|c|}{\bf Case} & \multicolumn{3}{c||}{\bf Running time
($s$)} & \#SMT & \#subformulas & \#states & \#transitions \\
\cline{3-5} \multicolumn{2}{|c|}{\bf study} & {\bf SMT encoding} & {\bf 
SMT solving} & {\bf Total} & variables  &  &  &\\
\hline\hline

\multirow{3}{*}{\TA} & $m=2$ & 5.43 & 0.31 & 5.74 & 8088 & 50654 & 
24 & 46 \\
\cline{2-5}	
& $m=4$ & 114  & 20 & 134 & 50460 & 368062 & 60 & 136 \\
\cline{2-5}
 & $m=6$ & 1721  & 865 & 2586 &  175728 & 1381118 & 112 & 274\\
\hline\hline
\multirow{3}{*}{\PW} & $m=2$ & 5.14 & 0.3 & 8.14 & 8088 & 43432 & 
24 & 46 \\
\cline{2-5}	
& $m=4$ & 207  & 40 & 247 & 68670 & 397852 & 70& 146 \\
\cline{2-5}
 & $m=6$ & 3980 & 1099 & 5079 & 274540 & 1641200 & 140 & 302\\
\hline\hline 
\multirow{4}{*}{\TS} & $h=(0, 1)$ & 0.83  & 0.07 & 0.9 & 1379 & 7913 & 
7 & 13 \\
\cline{2-5}
& $h=(0, 15)$ & 60 & 1607 & 1667 & 34335 & 251737 & 35 & 83 \\
\cline{2-5}
& $h=(4, 8)$ & 11.86 & 17.02 & 28.88 & 12369 & 87097 & 21 & 48 \\
\cline{2-5}
&  $h=(8, 15)$ & 60 & 1606 & 1666 & 34335 & 251737 & 35 & 83 \\
\hline\hline
\multirow{2}{*}{\PC} & s=(0) & 277 & 1996 & 2273 & 21220 & 1859004 & 20 
& 
158 \\
\cline{2-5}	
& s=(0,1) & 822 & 5808 & 6630 & 21220 & 5349205 & 20 & 280 \\
\cline{2-5}	
& s=(0,1,2) & 1690 & 58095 & 59785 & 21220 & 11006581 & 20 & 404 \\
\cline{2-5}	

\hline
    \end{tabular}}
{%
\vspace{2mm}
  \caption{Experimental results. {\bf TA:} Timing attack. {\bf PW:} Password 
leakage. {\bf TS:} Thread scheduling. {\bf PC:} Probabilistic conformance.} 
  \label{tab:results}
}
\end{table}

\section{Conclusion and Future Work}
\label{sec:concl}

We investigated the problem of specifying and model checking probabilistic 
hyperproperties of Markov decision processes (MDPs). Our study is motivated by 
the fact that many systems have probabilistic nature and are influenced by 
nondeterministic actions of their environment. We extended the temporal logic 
\HyperPCTL for DTMCs~\cite{ab18} to the context of MDPs by allowing formulas to 
quantify over schedulers. This additional expressive power leads to 
undecidability of the \HyperPCTL model checking problem on MDPs, but we
also showed that the undecidable fragment becomes decidable for 
non-probabilistic memoryless schedulers. Indeed, all applications 
discussed in this paper only require this type of schedulers.

Due to the high complexity of the problem, more efficient model checking 
algorithms are greatly needed. An orthogonal solution is to 
design less accurate and/or approximate algorithms such as statistical model 
checking that scale better and provide certain probabilistic guarantees about 
the correctness of verification. Another interesting direction is using 
counterexample-guided techniques to manage the size of the state space.

\bibliographystyle{splncs}
\bibliography{bibliography}

\iflong

\newpage
\appendix

\section{Proof of Theorem~\ref{thrm:undec}}

\noindent {\bf Theorem~\ref{thrm:undec}.} \ \thrmundec

\begin{proof}

We reduce the {\em emptiness} problem in {\em probabilistic B\"{u}chi automata} 
(PBA), which is known to be undecidable~\cite{bbg08}, to our problem.

\subsection{Probabilistic B\"{u}chi Automata}

PBA can be viewed as nondeterministic B\"{u}chi automata where the 
nondeterminism is 
resolved by a probabilistic choice. That is, for any state $q$ and letter $a$ 
in alphabet $\alphabet$, either $q$ does not have any $a$-successor or there 
is a probability distribution for the $a$-successors of $q$.

\begin{definition}
A {\em probabilistic B\"{u}chi automaton} (PBA) over a finite alphabet $\alphabet$ 
is a tuple $\pba= (Q, \delta, \alphabet, F)$, where $Q$ is a finite state 
space, $\delta: Q \times \alphabet \times Q \rightarrow [0,1]$ is the 
transition probability function, such that for all $q \in Q$ and $a \in 
\alphabet$:
$$\sum_{q' \in Q}(q, a, q') \in \{0, 1\}$$
%
%
%
and $F \subseteq Q$ is the set of {\em accepting states}. 
\end{definition}

A {\em run} for an infinite word $w=a_1a_2 \cdots \in \alphabet^\omega$ is an 
infinite sequence $\pi= q_0q_1q_2\cdots$ of states in $Q$, such that
$q_{i+1} \in \delta(q_i, a_{i+1}) = \{q' \mid \delta(q_i,a_{i+1}, q')>0\}$ for 
all $i\ \geq 0$. Let $\Inf(\pi)$ denote the set of states that are visited 
infinitely often in $\pi$. Run $\pi$ is called \emph{accepting} if 
$\Inf(\pi) \cap F \neq \emptyset$. Given an infinite input word $w \in 
\alphabet^\omega$, the behavior of $\pba$ is given by the infinite Markov chain 
that is obtained by unfolding $\pba$ into a tree using $w$. This is similar to 
an induced Markov chain from an MDP by a scheduler. Hence, standard concepts 
for Markov chains can be applied to define the acceptance probability of $w$ in 
$\pba$, denoted by $\Pr_\pba (w)$ or briefly $\Pr(w)$, by the probability 
measure of the set of accepting runs for $w$ in $\pba$. We define the accepted 
language of $\pba$ as:
$$\lang(\pba) = \{w \in \alphabet^\omega \mid \Pr_\pba(w)>0\}.$$
The {\em emptiness problem} is to decide whether or not $\lang(\pba) = \emptyset$ 
for a given input $\pba$.

\subsection{Mapping}

Our idea of mapping the emptiness problem in PBA to \HyperPCTL model checking 
for MDPs is as follows. We map a PBA to an MDP such that the words of the PBA 
are mimicked by the runs of the MDP. In other words, letters of the words in the 
PBA appear as propositions on states of the MDP. This way, the existence of a 
word in the language of the PBA corresponds to the existence of a scheduler 
that produces a satisfying computation tree in the induced Markov chain of the 
MDP.\\

\noindent \textbf{MDP $\mdp = (\states, \Act, \P, \AP, L)$:} \ Let $\pba= (Q, 
\delta, \alphabet, F)$ be a PBA with alphabet $\alphabet$. We obtain an MDP 
$\mdp = (\states, \Act, \P, \AP, L)$ as follows:

\begin{itemize}
 \item The set of states is $\states = Q \times \alphabet$.
 \item The set of actions is $\Act = \alphabet$.
 \item The transition probability function $\P$ is defined as follows:
 $$
 \P\Big((q, a), b, (q', a')\Big) =
 \begin{cases}
  \delta (q, b, q') ~~~~\text{ if } a' = b\\
  0 ~~~~~~~~~~~~~~~\text{otherwise}
 \end{cases}
$$

\item The set of atomic propositions is $\AP = \alphabet \cup \{f\}$, where $f 
\not \in \alphabet$ (we use $f$ to label the accepting states).

\item The labeling function $L$ is defined as follows. For each $a \in 
\alphabet$ and $q \in Q$, we have:
$$
 L(q, a) =
 \begin{cases}
  \{a,f\} ~~~~~~~~\text{ if } q \in F\\
  \{a\} ~~~~~~~~~~~~\text{otherwise}
 \end{cases}
$$

\end{itemize}

\noindent \textbf{HyperPCTL formula: } The \NHyperPCTL formula in our mapping 
is the following:
\begin{align*}
\varphi_\mathsf{map} = \exists \hscheduler.\exists 
\hstate(\hscheduler).\forall 
\hstate'(\hscheduler). &
\bigg(\pr\big(\G\bigwedge_{a \in 
\AP\setminus\{f\}}(a_{\hstate} \leftrightarrow 
a_{\hstate'})\big)=1\bigg) \; \wedge \\
& \bigg(\pr
    \Big(\F \pr
      \big(\G\pr
        (\F f_{\hstate}
        )=1
      \big)=1
    \Big) >0
  \bigg)
\end{align*}
Intuitively, the formula establishes connection between the PBA emptiness 
problem and \HyperPCTL model checking MDPs. In particular:

\begin{itemize}
 \item The existence of scheduler $\hscheduler$ in $\varphi_\mathsf{map}$ 
corresponds to the existence of a word $w$ in $\lang(\pba)$;

\item the state quantifiers and the left conjunct ensure that the path in 
the induced Markov chain and the PBA follow the sequence of actions 
(respectively, letters) in the witness to $\hscheduler$ (respectiely, $w$), and

\item the right conjunct mimics that a state in $F$ is visited with 
non-zero probability if and only if a state labeled by proposition $f$ is 
visited infinitely often in the MDP with non-zero probability.

\end{itemize}

\subsection{Reduction}

We now show that $\lang(\pba) \neq \emptyset$ if and only if $\mdp \models 
\varphi_\mathsf{map}$. We distinguish two cases:

\begin{itemize}
 \item ($\Rightarrow$) Suppose we have $\lang(\pba) \neq \emptyset$. This 
means there exists a word $w \in \alphabet^\omega$, such that $\Pr_\pba(w) > 
0$. We use $w$ to eliminate the existential scheduler quantifier and 
instantiate $\hscheduler$ in formula $\varphi_\mathsf{map}$. This induces a 
DTMC and now, we show that the induced DTMC satisfies the following \HyperPCTL 
formula as prescribed in~\cite{ab18}:
$$\exists \hstate.\forall \hstate'.
\bigg(\pr\big(\G\bigwedge_{a \in 
\AP\setminus\{f\}}(a_{\hstate} \leftrightarrow 
a_{\hstate'})\big)=1\bigg) \; \wedge 
 \bigg(\pr
    \Big(\F \pr
      \big(\G\pr
        (\F f_{\hstate}
        )=1
      \big)=1
    \Big) >0
  \bigg)
  $$
To this end, observe that the right conjunct is trivially satisfied due to the 
fact that $\Pr_\pba(w) > 0$. That is, since a state in $F$ is visited 
infinitely often with non-zero probability in $\pba$, a state labeled by $f$ 
in $\mdp$ is also visited infinitely often with non-zero probability. The left 
conjunct is also satisfied by construction of the mapped MDP, since the 
sequence of letters in $w$ appear in all paths of the induced DTMC as 
propositions.

 \item ($\Leftarrow$) The reverse direction is pretty similar. Since the 
answer to the model checking problem is affirmative, a witness to 
scheduler quantifier $\hscheduler$ exists. This scheduler induces a DTMC whose 
paths follow the same sequence of propositions. This sequence indeed provides 
us with the word $w$ for $\pba$. Finally, since the right conjunct in 
$\varphi_\mathsf{map}$ is satisfied by the MDP, we are guaranteed that $w$ 
reaches an accepting state in $F$ infinitely often with non-zero probability.
\end{itemize}

And this concludes the proof.\hfill\qed
\end{proof}

\section{Proof of Theorem~\ref{thrm:npc}}

\noindent {\bf Theorem.~\ref{thrm:npc}} \ \thrmnpc

\begin{proof}

 In order to show membership to \comp{NP}, let $\mdp$ be an MDP and $\varphi = 
\exists \hscheduler.\varphi'$ be a \NHyperPCTL formula. We show that given a 
solution to the problem, we can verify the solution in polynomial time. Observe 
that given a non-probabilistic memoryless scheduler as a witness to the 
existential quantifier $\exists \hscheduler$, one can compute the induced DTMC 
and then verify the DTMC against the resulting \HyperPCTL formula in polynomial 
time in the size of the induced DTMC~\cite{ab18}. 

Inspired by the proof technique introduced in~\cite{bf18}, for the lower bound, 
we reduce the SAT problem to our model checking problem.

\subsection{The Satisfiability Problem}

The SAT problem is as follows:

\begin{quote}
Let $\{x_1, x_2, \dots, x_n\}$ be a set of propositional variables. Given is a 
Boolean formula $y = y_1 \wedge y_2 \wedge \cdots \wedge y_m$, where each 
$y_j$, for $j \in [1, m]$, is a disjunction of at least three literals. Is 
$y$ satisfiable? That is, does there exist an assignment of truth values to 
$x_1, x_2, \dots, x_n$, such that $y$ evaluates to true?
\end{quote}

\subsection{Mapping}

We now present a mapping from an arbitrary instance of SAT to the model 
checking problem of an MDP and a \NHyperPCTL formula of the 
form $\exists \hscheduler. \exists \hstate(\hscheduler). \forall 
\hstate'(\hscheduler).\psi$. Then, we show that the MDP satisfies 
this formula if and only if the answer to the SAT problem is affirmative. 
Figure~\ref{fig:npc} shows an example.

\noindent \textbf{MDP $\mdp = (\states, \Act, \P, \AP, L)$: } 

\begin{itemize}
\item {\em (Atomic propositions $\AP$)} We include four atomic propositions: 
$p$ and $\bar{p}$ to mark the positive and negative literals in each clause 
and $c$ and $\bar{c}$ to mark paths that correspond to clauses of the SAT 
formula. Thus,
$$\AP = \big\{p, \bar{p}, c, \bar{c}\}.$$

\item {\em (Set of states $\states$)}  We now identify the members of $\states$:

\begin{itemize}

\item For each clause $y_j$, where $j \in [1, m]$, we include a state 
$r_j$, labeled by proposition $c$. We also include a state $r_0$ labeled by 
$\bar{c}$
  
\item For each clause $y_j$, where $j \in [1, m]$, we introduce the 
following $n$ states:
$$\Big\{v^j_i \mid i \in [1, n]\Big\}.$$
Each state $v^j_i$ is labeled with proposition $p$ if $x_i$ is a literal in 
$y_j$, or with $\bar{p}$ if $\neg x_i$ is a literal in $y_j$.

\item For each Boolean variable $x_i$, where $i \in [1, n]$, we include two 
states $s_i$ and $\bar{s}_i$. Each state $s_i$ (respectively, 
$\bar{s}_i$) is labeled by $p$ (respectively, $\bar{p}$).

\end{itemize}

\item {\em (Set of actions $\Act$)} The set of actions is $\Act = \{\alpha, 
\beta, \gamma\}$. Intuitively, the scheduler chooses action $\alpha$ 
(respectively, $\beta$) at a state $s_i$ or $\bar{s}_i$ to assign true 
(respectively, false) to variable $x_{i+1}$. Action $\gamma$ is the sole action 
available at all other states. 

\item {\em (Transition probability function $\P$)} We now identify the members 
of $P$. All transitions have probability 1, so we only discuss the actions.
 
\begin{itemize}
 
\item We add transitions $(r_j, \gamma, v^j_1)$ for each $j \in [1, m]$, where 
from $r_j$, the probability of reaching $v^j_1$ is 1. 

\item For each $i \in [1, n)$, we include four transitions $(s_i, \alpha, s_{i+1})$, 
$(s_i, \beta, \bar{s}_{i+1})$, $(\bar{s}_i, \alpha, s_{i+1})$, and 
$(\bar{s}_i, \beta, \bar{s}_{i+1})$. The intuition here is that when the 
scheduler chooses action $\alpha$ at state $s_i$ or $\bar{s}_i$, variable 
$x_{i+1}$ evaluates to true and when the scheduler chooses action $\beta$ at 
state $s_i$ or $\bar{s}_i$, variable $x_{i+1}$ evaluates to false in the SAT 
instance. We also include two transitions $(r_0, \alpha, s_1)$ and $(r_0, \beta,
\bar{s}_1)$ with the same intended meaning.

\item Finally, we include self-loops $(s_n, \gamma, s_n)$, $(\bar{s}_n, \gamma, 
\bar{s}_n)$, and $(v_n^j, \gamma, v_n^j)$, for each $j \in [1, m]$.

\end{itemize}

\end{itemize}

\begin{figure}[t]
\centering
\includegraphics[scale=1]{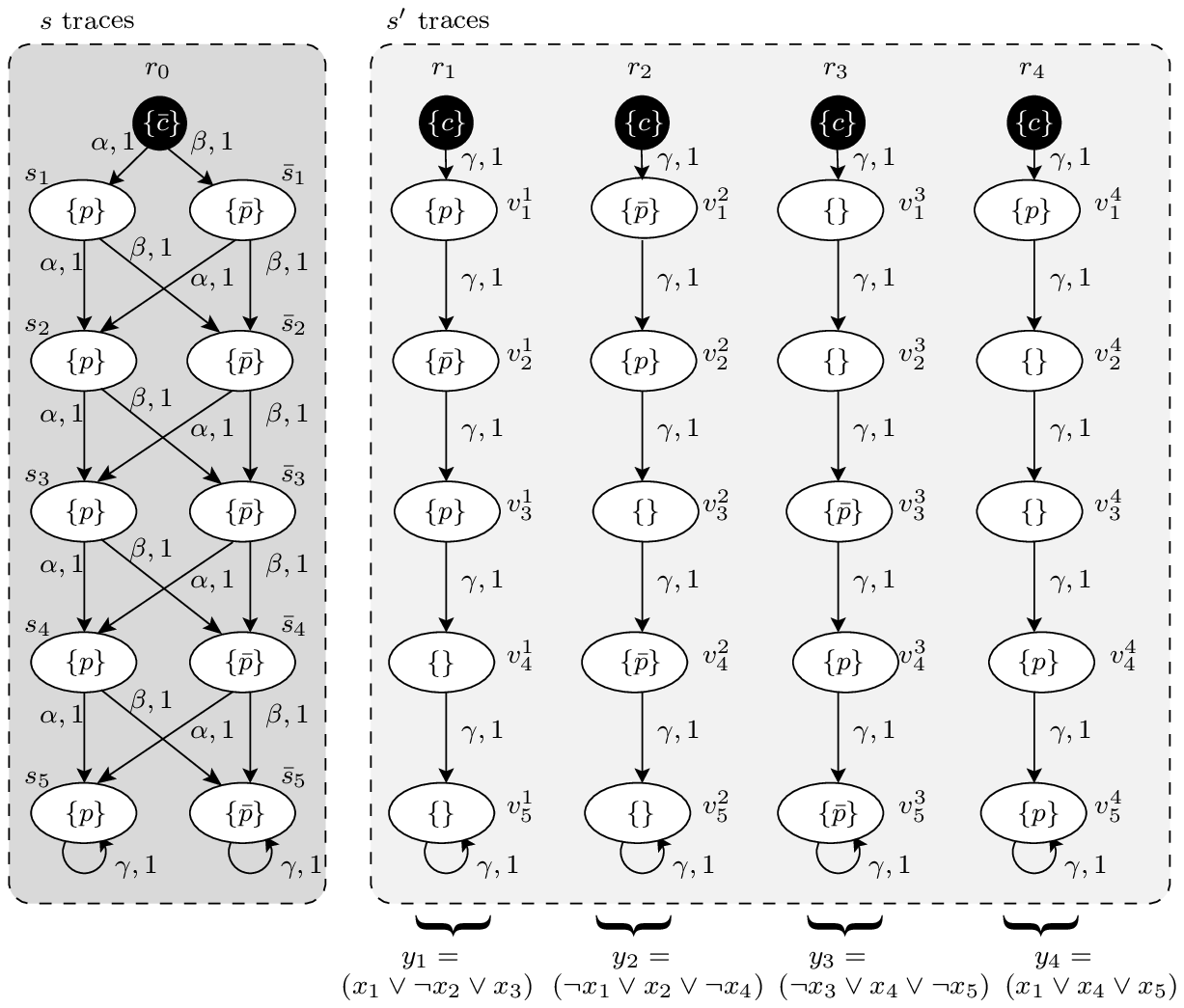}
\caption{Example of mapping SAT to \NHyperPCTL model checking.}
\label{fig:npc}
\end{figure}

\noindent \textbf{HyperPCTL formula: } The \NHyperPCTL formula in our mapping 
is the following:

\begin{eqnarray*}
  \varphi_{\map} &=&  \exists \hscheduler.\exists \hstate(\hscheduler).\forall 
\hstate'(\hscheduler). \bar{c}_{\hstate(\hscheduler)} \, \wedge \, 
\bigg(c_{\hstate'(\hscheduler)} \; \Rightarrow \\
&&\pr\bigg(\F 
\Big((p_{\hstate(\hscheduler)} \wedge p_{\hstate'(\hscheduler)}) 
\, \vee \, (\bar{p}_{\hstate(\hscheduler)} \wedge 
\bar{p}_{\hstate'(\hscheduler)})\Big)\bigg)=1\bigg)
\end{eqnarray*}

The intended meaning of the formula is that if there exists a scheduler that 
makes the formula true by choosing the $\alpha$ and $\beta$ actions, this 
scheduler gives us the assignment to the Boolean variables in the SAT instance. 
This is achieved by making all clauses true, hence, the $\forall 
\hstate'(\hscheduler)$ subformula.

\subsection{Reduction}

We now show that the given SAT formula is satisfiable if and only if the MDP 
obtained by our mapping satisfies the \NHyperPCTL formula $\varphi_\map$. 

\begin{description}

\item[($\Rightarrow$)] Suppose that $y$ is satisfiable. Then, there is an 
assignment that makes each clause $y_j$, where $j \in [1, m]$, true. We now use 
this assignment to instantiate a scheduler for the formula $\varphi_\map$. If 
$x_i = \tru$, then we instantiate scheduler $\hscheduler$ such that in state 
$s_{i-1}$ or $\bar{s}_{i-1}$, it chooses action $\alpha$. Likewise, if $x_i = 
\fals$, then we instantiate scheduler $\hscheduler$, such that in state 
$s_{i-1}$ or $\bar{s}_{i-1}$, it chooses action $\beta$. We now show that this 
scheduler instantiation evaluates formula $\varphi_\map$ to true. First 
observe that $\hstate(\hscheduler)$ can only be instantiate with state $r_0$ 
and $\hstate'(\hscheduler)$ can only be instantiate with states $r_j$, where 
$j \in [1,m]$. Otherwise, the left side of the implication in 
$\varphi_\map$ becomes false, making the formula vacuously true. Since each 
$y_j$ is true, there is at least one literal in $y_j$ that is true. If this 
literal is of the form $x_i$, 
then we have $x_i = \tru$ and the path that starts from $r_0$ will include 
$s_i$, which is labeled by $p$. Hence, the values of $p$, 
in both paths that start from $\hstate(\hscheduler)$ and 
$\hstate'(\hscheduler)$ are eventually equal. If the literal in $y_j$ is of 
the form $\neg x_i$, then $x_i = \fals$ and the path that starts from 
$\hstate(\hscheduler)$ will include $\bar{s}_i$. Again, the values of 
$\bar{p}$ are eventually equal. Finally, since all clauses are true, all 
paths that start from $\hstate'(\hscheduler)$ reach a state where the right 
side of the implication becomes true.

\item[($\Leftarrow$)] Suppose our mapped MDP satisfies formula $\varphi_\map$. 
This means that there exists a scheduler and state $\hstate(\hscheduler)$ 
that makes the subformula $\forall \hstate'(\hscheduler)$ true, i.e., the 
path that starts from $r_0$ results in making the inner \PCTL formula true for 
all paths that start from $r_j$. We obtain the truth assignment to the SAT 
problem as follows. If the scheduler chooses action $\alpha$ to state $s_i$, 
then we assign $x_i = \tru$. Likewise, if the scheduler chooses action $\beta$ 
to state $\bar{s}_i$, then we assign $x_i = \fals$. Observe that since in no 
state $p$ and $\bar{p}$ are simultaneously true and no path includes both $s_i$ 
and $\bar{s}_i$, variable $x_i$ will have only one truth value. Similar to the 
forward direction, it is straightforward to see that this valuation makes every 
clause $y_j$ of the SAT instance true.

\end{description}

And this concludes the proof.\hfill\qed
\end{proof}

\section{Proof of Theorem~\ref{thrm:sound}}
\label{sec:soundness}

\noindent {\bf Theorem.~\ref{thrm:sound}} \ \thrmsound

\begin{proof}
  The proof is by structural induction over the formula type. The cases for constants, atomic propositions, Boolean combinations and arithmetic expressions are straightforward. The remaining cases are the probabilistic temporal operators.

  \medskip
  
  For the next-operator $\pr(\Next\formula')$ (line \ref{line:semnext} in Algorithm \ref{algo:smt_sub}), we first encode the meaning of the operand $\formula'$ (line \ref{line:semnextB}); by induction assumption this encoding is sound.
  Since the operand is a state formula, its value is Boolean. The
  probability that the operand is true after one step is the sum of
  the probabilities to get to a state where the operand is true; we
  express this in pseudo-arithmetic by setting for each composed state
  $\vstate$ the value of
  $\ptruth{\vstate}{\formula'}$ to $1$ if $\formula'$ holds there and
  to $0$ otherwise. Using these pseudo-Boolean values, we express
  for each composed state $\vstate=(\state_1,\ldots,\state_n)$ and each scheduler $\vscheduler=(\scheduler_1,\ldots,\scheduler_n)$ the probability that $\formula'$ holds after one step by summing up
  for each possible successor $\vstate'=(\state_1',\ldots,\state_n')$ the
  probability $\Pi_{i=1}^n \P(\state_i,\alpha_i,\state_i')$ to get
  there in one step times the pseudo-Boolean value of $\ptruth{\vstate}{\formula'}$ (lines \ref{line:statechoice}--\ref{line:nextprob}).

  The semantics of unbounded until $\pr(\formula_1\U \formula_2)$ is
  encoded in Algorithm \ref{algo:smt_sub_unbounded_until}. Similarly
  to the next-operator, we recursively encode the truth of the
  Boolean-valued operands (line \ref{line:unboundedA}).  The remaining
  encoding follows for each composed state $\vstate$ and each
  scheduler $\vscheduler=(\scheduler_1,\ldots,\scheduler_n)$ the
  standard fixedpoint-encoding of the probability to satisfy the until
  formula in the induced DTMC \cite{bk08-book}.  This probability is
  $1$ for all states satisfying $\formula_2$ and $0$ for all states
  that do not satisfy any of the operands (line \ref{line:unboundedE}). Furthermore, the probability is
  also $0$ for all states from which no $\formula_2$-states are
  reachable; we assure this by requiring for all positive probabilities the existence of a finite loop-free path to a
  $\formula_2$-state using decreasing sequences of the 
  arithmetic variables $d_{\vstate,\formula_2}$ (line \ref{line:unboundedI}).
  For all other cases, the encoding is similar to the next-operator, summing up for all possible successor states the probabilities to get there times the probability to satisfy the until formula along paths starting from there (line \ref{line:unboundedH}).
  
  The case for unbounded until is a based on a technically rather complex case distinction, but it is just a direct encoding the semantics of bounded until.

\end{proof}

\fi

\end{document}